\documentclass{emulateapj}
\usepackage{natbib}
\usepackage{subfigure}

\newcommand{\captionfonts}{\small}

\makeatletter  
\long\def\@makecaption#1#2{%
  \vskip\abovecaptionskip
  \sbox\@tempboxa{{\captionfonts #1: #2}}%
  \ifdim \wd\@tempboxa >\hsize
    {\captionfonts #1: #2\par}
  \else
    \hbox to\hsize{\hfil\box\@tempboxa\hfil}%
    \fi
  \vskip\belowcaptionskip}
\makeatother   

\def\asec{$^{\prime\prime}$}
\def\lax{{$\mathrel{\hbox{\rlap{\hbox{\lower4pt\hbox{$\sim$}}}\hbox{$<$}}}$}}
\def\gax{{$\mathrel{\hbox{\rlap{\hbox{\lower4pt\hbox{$\sim$}}}\hbox{$>$}}}$}}

\begin{document}

\title{Two Populations of Molecular Clouds in the Antennae Galaxies}

\author{Lisa H. Wei\altaffilmark{1}, Eric Keto\altaffilmark{1}, and
  Luis C. Ho\altaffilmark{2}} \altaffiltext{1}{Harvard-Smithsonian
  Center for Astrophysics, 60 Garden Street, Cambridge, MA 02138}
\altaffiltext{2}{The Observatories of the Carnegie Institution for
  Science, 813 Santa Barbara Street, Pasadena, CA 91101}

\begin{abstract}

Super star clusters --- extremely massive clusters found predominately
in starburst environments --- are essential building blocks in the
formation of galaxies and thought to dominate star formation in the
high-redshift universe. However, the transformation from molecular gas
into these ultra-compact star clusters is not well understood. To
study this process, we used the Submillimeter Array and the Plateau de
Bure Interferometer to obtain high angular resolution ($\sim$~1\farcs5
or 160\,pc) images of the Antennae overlap region in CO(2--1) to
search for the molecular progenitors of the super star clusters. We
resolve the molecular gas distribution into a large number of clouds,
extending the differential cloud mass function down to a 5$\sigma$
completeness limit of 3.8$\times\,10^5\,M_{\odot}$. We identify a
distinct break in the mass function around log~$M_{\rm
  mol}/M_{\odot}\approx 6.5$, which separates the molecular clouds
into two distinct populations. The smaller, less massive clouds reside
in more quiescent areas in the region, while the larger, more massive
clouds cluster around regions of intense star formation. A broken
power-law fit to the mass function yields slopes of $\alpha =
-1.39\pm\,0.10$ and $\alpha = -1.44\pm\,0.14$ for the low- and
high-mass cloud population, well-matched to the mass function found
for super star clusters in the Antennae galaxies. We find large
velocity gradients and velocity dispersions at the locations of
intense star formation, suggestive of compressive shocks. It is likely
that these environmental factors contribute to the formation of the
observed massive molecular clouds and super star clusters in the
Antennae galaxies.

\end{abstract}


\section{Introduction}\label{section.intro}

In areas of lower extinction between the dust lanes of starbursts,
observations by the {\it Hubble Space Telescope} (\textit{HST}) have
revealed an important connection between the starburst phenomenon and
a class of objects known as ``super star clusters'' (SSCs;
\citealt{larsen10,zwart10}). SSCs represent the most extreme mode of
star formation known. Aside from their extraordinary luminosities,
which range from 1--100 times that of the R136 cluster of 30 Doradus
in the Large Magellanic Cloud \citep{holtzman92,whitmore03,larsen04},
these clusters are exceptionally compact (radii $\leq 1 - 5$~pc;
\citealt{larsen10,zwart10}) and young ($\lesssim$ few -- 50 Myr;
\citealt{zwart10}). Estimates based on high-resolution spectroscopy
and population synthesis models suggest that cluster masses range from
10$^4$ to 10$^6$\,$M_{\odot}$ (e.g.,
\citealt{ho96a,ho96b,zhang99,mengel02,larsen04,melo05,mccrady07}). SSCs
comprise a significant fraction of the UV, H$\alpha$, and $B$-band
light emitted by young stars in starburst regions
\citep{barth95,meurer95,maoz96,zepf99,fall05}. This suggests that in
these regions SSCs are the basic building blocks of star formation
\citep{ho97}.

The similarity in size and mass of the SSCs and globular clusters
suggests that the SSCs might be present-day analogs of young globular
clusters \citep{holtzman92,mccrady07}. Previously thought to form
strictly in the primordial epoch of galaxy evolution, globular
clusters may in fact be forming in the current epoch as SSCs in
starburst environments \citep{ashman01}.

The formation of SSCs, however, is not well understood. Some key
questions include: How is the mass of the precursor molecular cloud,
equivalent to a giant molecular cloud (GMC), converted to stars
quickly and efficiently enough for the star cluster to remain bound?
How are the conditions in starbursts different from ``ordinary'' star
formation?  Are there other distinguishing factors that differentiate
progenitor clouds of SSCs from normal GMCs?

Recent studies suggest that it is high environmental pressure that
increases the gas density and ultimately leads to the formation of
massive star clusters (e.g., \citealt{ashman01,herrera11}). One
hypothesis suggests that a sudden galactic-scale increase in external
pressure, possibly from shock compression associated with mergers,
crushes entire GMCs rapidly and efficiently, forming individual SSCs
\citep{jog92,ashman01}. CO(2--1) line observations of M82 with the
Owens Valley Millimeter Array (OVRO) with 17\,pc spatial resolution by
\citet{keto05} support this, showing that the mass spectrum of GMCs in
the starburst region is indistinguishable from the upper range of the
mass spectrum of SSCs in M82 \citep{melo05}. These observations
further show that the clouds with active star formation are in
supersonic compression, with both lateral and line-of-sight velocity
gradients within the individual clouds.

In this hypothesis, the compression itself is created by the rapid
ionization of the interstellar medium (ISM) surrounding the progenitor
GMCs. The initial ionization may be caused by the dissipation of
kinetic energy in the diffuse ISM of colliding galaxies. The burst of
massive star formation that follows then releases enough energy to
further propagate the ionization and compress nearby molecular
clouds. This hypothesis for ionization-driven implosion draws a
fascinating connection between the present-day SSCs and the ancient
globular clusters in that the globular clusters may also have formed
in clouds compressed by rapid ionization in the epoch of reionization
in the primordial universe at $z \approx 6$
\citep{cen01,vandenbergh01}.

Alternatively, SSCs may form in the high-pressure environment of the
centers of super-giant molecular complexes ($M \ge 10^7$\,$M_{\odot}$;
\citealt{harris94,wilson03}). In this hypothesis, cores that are the
size of GMCs, but are within larger molecular complexes, are slowly
compressed (on a crossing time) by the pressure of the overlying
molecular gas. Cluster formation proceeds as a scaled-up version of
normal star formation in our own Galaxy \citep{elmegreen97}. In
support of this alternative, CO(1--0) line observations of the
Antennae galaxies show huge concentrations of molecular gas that could
be interpreted as super-giant molecular complexes
\citep{wilson03}. Yet because the angular resolution of these
observations (3\farcs2$\times$4\farcs9; $\sim$\,420\,pc at the adopted
distance of the 22\,Mpc) is insufficient to resolve individual GMCs,
the question remains whether the appearance of the extraordinary
massive clouds is simply the result of limited spatial resolution.

In this paper we revisit the Antennae galaxies, NGC~4038 and NGC~4039,
with higher resolution imaging of the molecular gas to study the
formation of SSCs.  As one of the most spectacular starbursts in the
local universe, the Antennae has been studied in great detail over the
last several decades in various wavelengths, ranging from the X-rays
to the mm/cm regime, and everything in between (e.g.,
\citealt{mirabel98,neff00,wilson00,wilson03,fabbiano01,fabbiano03,hibbard01,hibbard05,zhang01,kilgard02,mengel02,zazas02a,zazas02b,fall05}). It
is also one of the best-studied SSC populations
\citep{whitmore95,whitmore99,whitmore10}.

The overlap region of the Antennae, where NGC~4038 and NGC~4039
collide, is one of the most intensely and violently star-forming
regions in the nearby universe
\citep{evans97,mirabel98,gao01,gao08,brandl09,klaas10,zhang10}. The
star formation rate (SFR) for the entire NGC~4038/39 system is high,
estimated to be between 6.6--20 $M_{\odot}$\,yr$^{-1}$
\citep{zhang01,brandl09}. However, most of the star formation occurs
in the overlap region, with SFRs of 0.7--2.0\,$M_{\odot}$\,yr$^{-1}$
for individual star-forming regions on size scales of
500$\times$1200\,pc$^2$ \citep{brandl09}. Recent studies also find
intriguing evidence for shock-driven star formation \citep{gao01,
  zhang10}, in agreement with the \citeauthor{jog92} model. Detailed
population synthesis analysis of the clusters in these regions finds
the median age of the clusters to be some of the youngest in the
system, $\lesssim$10\,Myr in many cases (e.g.,
\citealt{snijders07,zhang10}).

In this paper, Section \ref{section.intro} comprises this
introduction. In Section \ref{section.data} we present new,
high-resolution CO(2--1) data from the Submillimeter Array (SMA;
\citealt{ho04}) and the IRAM Plateau de Bure
Interferometer\footnote{IRAM is supported by INSU/CNRS (France), MPG
  (Germany) and IGN (Spain).} (PdBI). Section \ref{section.clumps}
presents some basic properties of the molecular gas in the observed
region and discusses the implications. In Section \ref{section.sf} we
compare the distribution of molecular gas mass, velocity, and velocity
dispersion with star formation tracers. In Section
\ref{section.summary}, we summarize our results.  Throughout this
paper we adopt a distance of 22\,Mpc for the Antennae
\citep{schweizer08}.

\begin{figure}
\hspace{.25cm}
\includegraphics[scale=.45,angle=0]{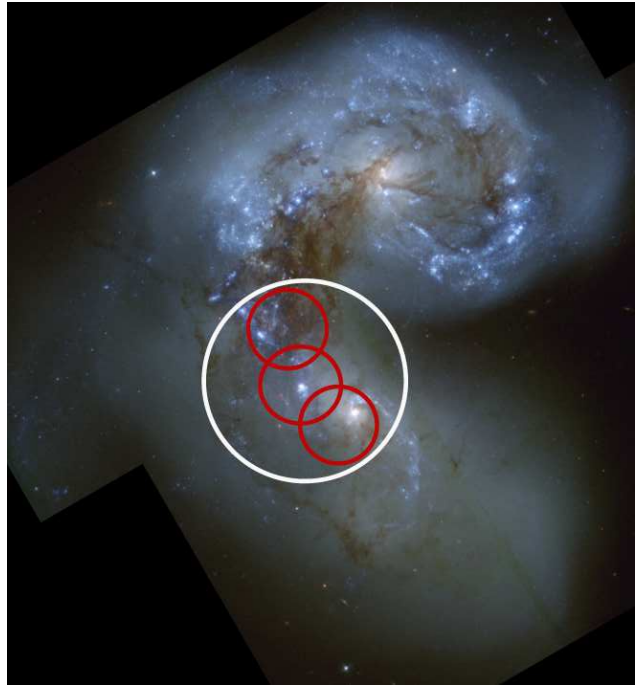}
\caption{Three-color image of the Antennae from the {\it HST} Advanced
  Camera of Surveys (ACS), using data taken by
  \citet{whitmore10}. Filters used to create the image are F435W
  (blue), F550M (green), and F814W (red).  The image has dimensions of
  160\arcsec$\times$195\arcsec, and it is displayed using an arcsinh
  stretch \citep{lupton04}. The white circle shows the primary beam of
  our SMA observations, and red circles show our PdBI pointings. See
  Table \ref{table.observations} for more
  information. \label{fig.hst}}
\end{figure}

\section{Observations}\label{section.data}

\subsection{The Submillimeter Array}

The SMA observations were taken in May 2008 and February 2011 in the
compact and extended array configurations, respectively, with a single
pointing with a phase center ($\alpha$,$\delta$)$_{\rm J2000}$ =
(12$^{\rm h}$01$^{\rm m}$54$^{\rm s}$.70,
$-$18$^{\circ}$53$^{\prime}$05\farcs0). Figure \ref{fig.hst} shows our
spatial coverage of the overlap region\footnote{Based on observations
  made with the NASA/ESA Hubble Space Telescope, obtained from the
  data archive at the Space Telescope Science Institute. STScI is
  operated by the Association of Universities for Research in
  Astronomy, Inc. under NASA contract NAS 5-26555.}. At 230\,GHz, the
primary beam of the SMA covers $\sim$54$^{\prime\prime}$. Baselines
ranged from 16 to 139\,meters (compact array) and 44 and 226\,meters
(extended array). Typical system temperatures ranged between 100 and
300\,K for both observations. The correlator was tuned to the
frequency of CO(2--1) at 230.53799\,GHz, with Doppler tracking
accounting for the recessional velocity of the Antennae galaxies. For
the May 2008 track, the correlator was configured to cover a 2\,GHz
bandwidth with 24 overlapping windows each 104\,MHz in width, with a
spectral resolution of 0.4\,MHz ($\sim$0.5\,km\,s$^{-1}$
channels). The February 2011 track was configured slightly differently
due to the bandwidth doubling upgrade by the SMA in 2009, covering
4\,GHz with 48 overlapping windows each 104\,MHz in width, with a
spectral resolution of 0.8\,MHz ($\sim$1\,km\,s$^{-1}$ channels).

The SMA data were reduced using the MIR data reduction package
developed at OVRO and the SMA. Data were flagged for bad channels,
antennas, weather, and pointing. Bandpass calibration was done on a
bright quasar, 3C~454.3 for the May 2008 observation and 3C~84 for the
February 2011 observation. We performed atmospheric phase calibrations
with 1127$-$189 and 1256$-$057 every 10 minutes. Absolute flux
calibration was done using Callisto and Titan. The channels were
Hanning smoothed and averaged to achieve a final resolution of
4.9\,km\,s$^{-1}$.

\subsection{Plateau de Bure Interferometer}

The PdBI observations were taken between February and April of 2006 in
the B and C arrays, respectively. Because of the smaller
($\sim$21$^{\prime\prime}$) primary beam of the PdBI 15\,meter
antennas, three separate pointings were obtained to cover
approximately the same area of the overlap region as the SMA primary
beam at 230\,GHz (Figure \ref{fig.hst} \& Table
\ref{table.observations}). Each pointing was observed with a single
$\sim$7 hour track at the PdBI. Baselines ranged from 24 to 82\,meters
in the C array and 40 to 280\,meters in the B array. Typical system
temperatures ranged from 300 to 1000\,K due to the low elevation of
the Antennae for the PdBI. The correlator was configured to cover the
CO(2--1) line with four windows each 160\,MHz in width, with 160
channels in each window with spectral resolution of 1.25\,MHz
($\sim$1.6\,km\,s$^{-1}$ channels).

The PdBI data were reduced using the GILDAS CLIC package. Bandpass and
absolute flux calibration were done using 3C~273, and the atmospheric
phase was calibrated using 1124$-$018 every 20 minutes. The channels
were Hanning smoothed and averaged to achieve a final resolution of
4.9\,km\,s$^{-1}$.

\subsection{Imaging}\label{section.imaging}

To recover the more extended flux from the shorter spacings and
increase the signal-to-noise ratio, we combine the high-resolution SMA
and PdBI data together with the SMA compact array observations in the
($u$,$v$) plane. After reduction, the PdBI ($u$,$v$) data were saved
as FITS files, then converted into the MIRIAD format using the task
{\tt fits}. Header information including telescope name, longitude,
latitude, and channel resolution were manually put into the PdBI
MIRIAD datasets. We combined the data with the MIRIAD software package
\citep{sault95}, using natural weighting and an additional weighting
in inverse proportion to the noise as estimated by the system
temperature. The data cube was cleaned to a cutoff of 2\,$\sigma_{\rm
  chan}$ in the residual image. The synthesized beam of the combined
SMA$+$PdBI image is 3\farcs3$\times$1\farcs5, with a position angle of
41$^{\circ}$. The rms in a single 4.9\,km\,s$^{-1}$ channel
($\sigma_{\rm chan}$) is 30\,mJy\,beam$^{-1}$.

\begin{figure*}
\vspace{-3cm}
\includegraphics[scale=.9,angle=0]{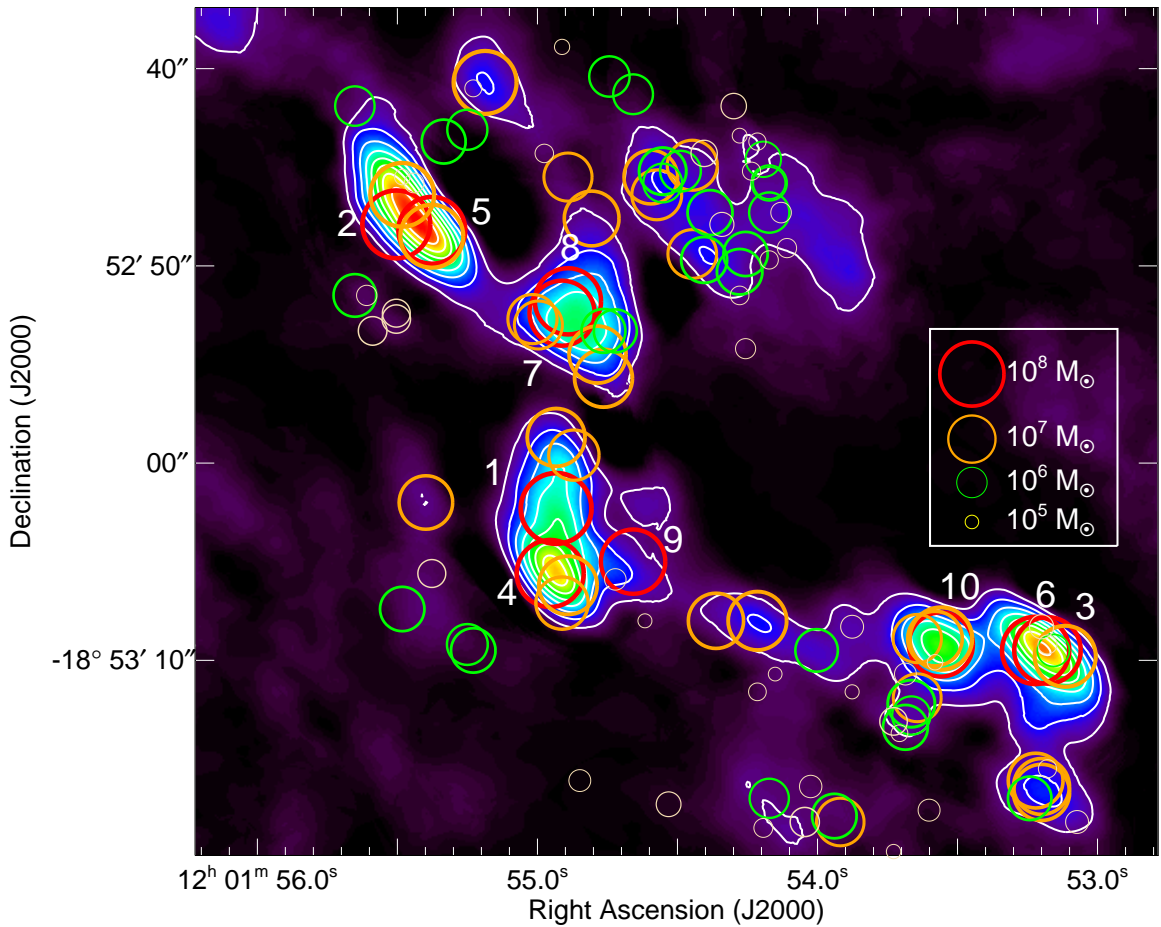}
\vspace{-3.cm}
\caption{Velocity-integrated CO(2--1) emission from the combined
  SMA$+$PdBI data of the overlap region of the Antennae. Contours
  start at 3 times the rms of the image ($\sigma$ = 3.71
  Jy\,beam\,$^{-1}$\,km\,s$^{-1}$) and increase by
  3\,$\sigma$. Intensities range from 0--100
  Jy\,beam\,$^{-1}$\,km\,s$^{-1}$. The synthesized beam of this image
  is 3\farcs3$\times$1\farcs5, corresponding to a physical scale of
  $374\times160$\,pc. We denote the locations of the clouds found by
  {\tt clumpfind} with circles. Symbol size, thickness, and color
  reflect the molecular gas mass of the cloud estimated from the CO
  brightness. We find multiple overlapping clouds, with the most
  massive clouds concentrated at the peaks of integrated CO emission
  and less massive clouds in the surrounding regions. Numbers indicate
  the clouds whose intensity-weighted velocity is shown in Figure
  \ref{fig.clumps_velo}. We discuss this figure in further detail in
  \S~\ref{section.clumps}. \label{fig.big_clumps}}
\end{figure*}

Velocity-integrated emission (zeroth moment), intensity-weighted
velocity (first moment), and intensity-weighted dispersion (second
moment) maps were made from a masked data cube. The masks were made by
smoothing the original cube to a resolution of 6$^{\prime\prime}$ and
masking emission that was less than 2$\sigma$ or did not cover two
contiguous channels. Figure \ref{fig.big_clumps} shows the
velocity-integrated CO(2--1) emission of our combined PdBI$+$SMA
observations of the overlap region.

\begin{deluxetable}{ll}
  \tabletypesize{\footnotesize}
  \tablewidth{0pt}
  \tablecaption{Telescope Pointing Centers\label{table.observations}}

  \tablehead{ 
    \colhead{Position} & 
    \colhead{Phase Center} 
      \\ 
    \colhead{} & 
    \colhead{($\alpha_{2000},\delta_{2000}$)}  
  }

  \startdata
  SMA Comp  & 12$^{\rm h}$01$^{\rm m}$54$^{\rm s}$.70,$-$18$^{\circ}$53$^{\prime}$05\farcs0  \\
  SMA Ext   & 12$^{\rm h}$01$^{\rm m}$54$^{\rm s}$.70,$-$18$^{\circ}$53$^{\prime}$05\farcs0  \\
  PdBI A    & 12$^{\rm h}$01$^{\rm m}$55$^{\rm s}$.00,$-$18$^{\circ}$52$^{\prime}$50\farcs0  \\
  PdBI B    & 12$^{\rm h}$01$^{\rm m}$54$^{\rm s}$.70,$-$18$^{\circ}$53$^{\prime}$05\farcs0  \\
  PdBI C    & 12$^{\rm h}$01$^{\rm m}$54$^{\rm s}$.00,$-$18$^{\circ}$53$^{\prime}$15\farcs0 
  \enddata

\end{deluxetable}

\section{Properties of the Molecular Gas}\label{section.clumps}

\subsection{Finding Clouds}\label{section.clumpfind}
To extract individual clouds from our data cubes, we utilize the
algorithm {\tt clumpfind} by \citet{williams94} on our data cube where
the PdBI and SMA primary beams overlap. {\tt clumpfind} contours the
input data cube in steps that are integer multiples of the input
channel rms and identifies clouds as emission peaks separated by more
than one contour. The extent of each cloud is determined by following
each peak down to the first (lowest) contour. Multi-peaked emission is
separated by a ``friends-of-friends'' algorithm where pixels adjacent
to the peak and subsequent adjacent pixels are identified as related.

The default contour and detection thresholds recommended for {\tt
  clumpfind} by \citet{williams94} are 2\,$\sigma_{\rm chan}$. We
tested three different detection thresholds (2, 2.5, and
3\,$\sigma_{\rm chan}$ while holding the contour threshold constant at
2\,$\sigma_{\rm chan}$). However, we find that most of the results do
not depend strongly on the choice of detection threshold. We discuss
this in more detail with the individual results. We set the {\tt
  clumpfind} search parameters to contour at twice the channel rms
(2\,$\sigma_{\rm chan}$) with the detection threshold at
3\,$\sigma_{\rm chan}$ to detect the faintest clouds with a higher
confidence level and better avoid spurious detections of low-mass
clouds.

\subsection{Fluxes and Mass Estimates}\label{section.massflux}

We estimate the mass of each cloud from its CO luminosity following 
\citet{solomon92}:

\begin{equation}
\frac{L_{\rm CO}}{\rm K\,km\,s^{-1}\,pc^2} = 3.25\times10^{7}\nu^{-2}_{\rm rest}(1+z)^{-1}\left(\frac{D_L}{\rm Mpc}\right)^2\left(\frac{F^{\prime}_{\rm CO}}{\rm Jy\,km\,s^{-1}}\right),
\end{equation}

\noindent 
where $F^{\prime}_{\rm CO}$ is the velocity-integrated flux, $\nu_{\rm
  rest}$ is the rest frequency of the line in GHz [230.538\,GHz for
  CO(2--1)], $z$ is the redshift, and $D_L$ is the luminosity
distance, which we fix to 22\,Mpc \citep{schweizer08}.  The CO
luminosity is related to the molecular gas mass by

\begin{equation}
M_{\rm H_2} = \alpha_{\rm CO}\,L_{\rm CO},
\end{equation}

\noindent where $\alpha_{\rm CO} = 2 m_{\rm H} X_{\rm CO}$. 

There has been some debate over the value of $X_{\rm CO}$ in the
Antennae. \citet{zhu03} find $X_{\rm CO}$ to be 5--13 times smaller
than the Galactic value ($X_{\rm CO, gal}\,=2.8\times$10$^{20}$
(K\,km\,s$^{-1}$)$^{-1}$). Analysis of resolved CO(1--0) clouds by
\citet{wilson03}, however, suggest that $X_{\rm CO}$ in the Antennae
is much larger, with an average value of 1.3$\times$10$^{20}$
(K\,km\,s$^{-1}$)$^{-1}$ and within the errors of the Galactic
value. More recently, models of CO and [C~II] data by \citet{schulz07}
suggest that $X_{\rm CO}$ does not deviate much from the Galactic
value. Based on these results and for consistency with
\citet{wilson03}, we adopt $\alpha_{\rm CO} = 4.8\,M_{\odot}$
(K\,km\,s$^{-1}$\,pc$^2$)$^{-1}$, which corresponds to a CO--to--H$_2$
conversion factor $X_{\rm CO}$ = 3$\times$10$^{20}$
(K\,km\,s$^{-1}$)$^{-1}$.

We also assume a line intensity ratio of $I_{\rm CO(2-1)}$/$I_{\rm
  CO(1-0)} =$ 1.0. \citet{zhu03} find the $I_{\rm CO(2-1)}$/$I_{\rm
  CO(1-0)}$ line intensity ratio to range from 1.0 to 1.2 in the
overlap region of the Antennae with their single-dish
observations. This is also the typical value found for nearby disk
galaxies, as well as the central region of the Milky Way
\citep{braine93,sawada01}. We multiply $M_{\rm H_2}$ by 1.36 to
account for heavier elements to obtain the total molecular gas mass,
$M_{\rm mol}$.

One concern, especially with the higher resolution imaging, is the
problem of missing flux from the lack of short spacings in
interferometric observations. In order to estimate the percentage of
the total flux captured by our interferometric observations, we first
estimate the total flux in the region covered by other
observations. CO(1--0) single-dish mapping of the Antennae with the
NRAO 12\,m by \citet{gao01} gives a total flux of
3172\,Jy\,km\,s$^{-1}$. Note that the region mapped by \citet{gao01}
is much larger than the region covered by our observations or that of
\citet{wilson03}. They used the single-dish spectra of \citet{gao01}
to estimate a total CO(1--0) flux of 1654\,Jy\,km\,s$^{-1}$ in their
smaller area mapped by the OVRO interferometers. We can use the
estimate of \citet{wilson03} to determine the total CO(1--0) flux in
the area mapped by our observations. Since our observations do not
include NGC~4038, we subtract 256\,Jy\,km\,s$^{-1}$ from Wilson's
estimate to obtain a total single-dish CO(1--0) flux for our region of
1398\,Jy\,km\,s$^{-1}$.

We can compare the predicted total CO flux with the flux measured in
our interferometric observations. We determine the flux by summing up
the fluxes of the clouds found by {\tt clumpfind} in our combined
SMA$+$PdBI data. We obtain a total flux of 1358\,Jy\,km\,s$^{-1}$,
which is very close to the single-dish flux assuming a line intensity
ratio of $I_{\rm CO(2-1)}$/$I_{\rm CO(1-0)} =$ 1.0. If we assume a
line intensity ratio of $I_{\rm CO(2-1)}$/$I_{\rm CO(1-0)} =$ 1.1
\citep{zhu03}, then we are missing only $\sim$12\% of the total
flux. This indicates that most of the CO mass may be in discrete
clouds rather than distributed in the diffuse ISM in the overlap
region. Note that this may not be the case for regular star-forming
disks, where the gas tends to be more diffuse. Lower cutoff thresholds
in {\tt clumpfind} yield total fluxes of 1472 and
1614\,Jy\,km\,s$^{-1}$ for the 2.5 and 2$\sigma_{\rm chan}$
thresholds, respectively, which imply that noise clumps may be picked
up as real clouds at these threshold levels.

\subsection{Basic Cloud Properties: Two Populations}\label{section.properties}

With the {\tt clumpfind} parameters adopted in
\S~\ref{section.clumpfind}, we detect a total of 132 clouds. We list
all the clouds and their basic properties in Table
\ref{table.clumps}. As expected, if we use the lower detection
thresholds, {\tt clumpfind} extracts a larger number of clouds (total
of 332 and 224 clouds for the 2 and 2.5$\sigma_{\rm chan}$ thresholds,
respectively). Most of the new clouds detected with the lower
thresholds are low mass and unresolved --- the number of clouds with
log~$M_{\rm mol}/M_{\odot} <$ 6.5 doubles and triples with the 2.5 and
2$\sigma_{\rm chan}$ thresholds and therefore are not significant in
our analysis.

\begin{figure*}
\hspace{-1cm}
\includegraphics[scale=.75,angle=0]{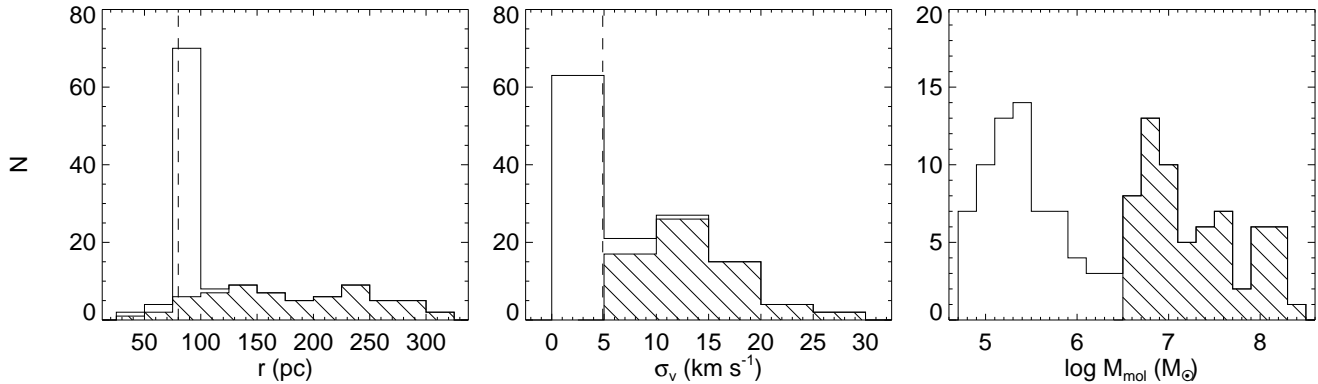}
\caption{Histogram of radius, velocity dispersion, and molecular gas
  mass for the clouds found by {\tt clumpfind}. Clouds that were
  unresolved spatially or in velocity are assigned upper limits of
  80\,pc and 4.9\,km\,s$^{-1}$, respectively, which we mark in the
  first two panels with dashed vertical lines. The clouds are broken
  up into two groups, above and below log~$M_{\rm mol}/M_{\odot} =
  6.5$ (shaded and open histogram, respectively), which corresponds to
  a distinct break in the differential mass function discussed in
  \S~\ref{section.massspec}.\label{fig.threepanel}}
\end{figure*}

We find a large range of values for the basic properties of the 132
clouds (Figure \ref{fig.threepanel} and Table 3).  The molecular
masses (discussed in \S~\ref{section.massflux}) span over 3 orders of
magnitude in dynamic range, from $M_{\rm mol} \approx 10^{5}$ to
$2.4\times10^{8}\,M_{\odot}$, with a mean of
$2.0\times10^{7}\,M_{\odot}$, a median of $3.0\times10^{6}\,
M_{\odot}$, and standard deviation (calculated in log space) of 1.0
dex. One of the most interesting characteristics of the mass
distribution is that it appears distinctly bimodal, with a maximum
around our sensitivity limit of $M_{\rm mol} \approx 10^{5}
\,M_{\odot}$ and clear secondary maximum near $M_{\rm mol} \approx
10^{7} \,M_{\odot}$. The exact location of the minimum is somewhat
arbitrary, within the range of log~$M_{\rm mol}/M_{\odot}\approx
6.3-6.7$. From considerations of the mass function
(\S~\ref{section.massspec}), we pick a fiducial value of $\log M_{\rm
  mol}/M_{\odot} = 6.5$ to separate the two populations.  With this
choice, there are 68 clouds in the low-mass group and 64 clouds in the
high-mass group, whose respective median masses are $M_{\rm mol} =
2.3\times10^{5}\,M_{\odot}$ and $M_{\rm mol} =
1.4\times10^{7}\,M_{\odot}$.

Values of radius and velocity dispersion estimated by {\tt clumpfind}
are corrected for spatial and spectral resolution following equation
A7 from \citet{williams94}. We find that the radii of the clouds span
a very large range, from $r = 35$ to 317\,pc.  Of the 132 clouds
found, 64 are unresolved spatially. We assign an upper limit of 80\,pc
to the radius of these clouds, half of our minimum resolution diameter
of 160\,pc.  Using the Kaplan-Meier product-limit estimator
\citep{feigelson85}, which properly accounts for the upper limits, we
find a mean radius of 119.2$\pm$7.3\,pc and a median value of 77.0\,pc
for the entire sample.  The distribution of sizes, like that of mass,
appears bimodal.  In fact, given the large fraction of upper limits at
the small-$r$ end of the distribution, the true separation between the
two peaks must be even greater than it appears.  Splitting the sample
into the low-mass and high-mass groups defined above, the respective
mean and median radii become $r$ = (57.0$\pm$7.1, 48.0) pc and
(180.6$\pm$8.8, 170.0) pc.  In light of the large number of upper
limits, the statistics for the low-mass subsample should be treated
with caution.

The clouds on the upper end of the size distribution are larger than
any known GMCs in the Milky Way \citep{solomon87,heyer09} or in nearby
disk galaxies \citep{bolatto08}.  While blending may be an issue (but
see arguments below as to why we do not think this is a major
problem), at face value the largest clouds represent a distinct
class. \citet{wilson00} described these very large clouds as
super-giant molecular complexes. Our spatial resolution of 80\,pc is
high enough to resolve the most massive clouds into smaller clouds,
but nonetheless they appear to be single entities. There is very good
correspondence between cloud size and cloud mass; the most massive
clouds are also the largest in physical extent.  The corollary
implication, therefore, is that the most massive clouds in our sample
are intrinsically as massive as they appear.  Their exceptionally high
masses are not a spurious artifact of insufficient resolution.  This
is one of the central results of this study.

Of the 132 clouds found, 86 are resolved in velocity, with velocity
dispersions ($\sigma_v$) ranging from 0.8--25.6\,km\,s$^{-1}$.  For
the unresolved clouds, we assign an upper limit of 4.9\,km\,s$^{-1}$,
the channel size of our data cube.  There is an overlap of 45 clouds
that have limits in both radius and velocity dispersion. Note that a
few clouds have radius and/or $\sigma_v$ smaller than the upper limit
because they were bright enough to be resolved.  Unlike the sizes and
masses, the distribution of velocity dispersions is not as cleanly
separated between the high-mass and low-mass groups, although lower
mass clouds tend to have lower values of $\sigma_v$.  We calculate a
mean and median velocity dispersion of 7.8$\pm$0.6\,km\,s$^{-1}$ and
6.2 km\,s$^{-1}$, respectively, for the sample as a whole.  Considered
separately, the low-mass and high-mass groups have mean and median
dispersions of $\sigma_v$ = (2.6$\pm$0.3, 2.0) km\,s$^{-1}$ and
(13.3$\pm$0.6, 12.0) km\,s$^{-1}$, respectively.  Again, the
statistics of the low-mass group may be uncertain because of the large
fraction (68\%) of upper limits.

\begin{deluxetable*}{l|rrr|rrr|rrr|ll}
\tablenum{3}
  \tabletypesize{\footnotesize}
  \tablewidth{0pt}
  \tablecaption{Cloud Statistics}
 
  \tablehead{ 
    \colhead{} & 
    \colhead{} & 
    \colhead{$r$} & 
    \colhead{} & 
    \colhead{} & 
    \colhead{$\sigma_v$} & 
    \colhead{} &
    \colhead{} &
    \colhead{$\log M_{\rm mol}$} &
    \colhead{} & 
    \colhead{$\alpha$} & 
    \colhead{$\beta$} 
    \\
    \colhead{} & 
    \colhead{Mean} &
    \colhead{Error} &
    \colhead{Median} & 
    \colhead{Mean} &
    \colhead{Error} &
    \colhead{Median} & 
    \colhead{Mean} & 
    \colhead{StDev} &
    \colhead{Median} & 
    \colhead{} & 
    \colhead{}
    \\
    \colhead{} & 
    \colhead{} & 
    \colhead{(pc)} & 
    \colhead{} & 
    \colhead{} & 
    \colhead{(km\,s$^{-1}$)} & 
    \colhead{} &
    \colhead{} &
    \colhead{($\log M_{\odot}$)} &
    \colhead{} & 
    \colhead{} & 
    \colhead{} 
  }
  \startdata
  All Clouds                          & 119.2 & 7.3 &  77.0 &  7.8 & 0.6 &  6.2 & 7.3 & 1.0 & 6.5 &                   &                \\
  log $M_{\rm mol}/M_{\odot} \geq  6.5$ & 180.6 & 8.8 & 170.0 & 13.3 & 0.6 & 12.0 & 7.6 & 0.5 & 7.2 &  $-$1.44$\pm$0.14 &   4.01$\pm$1.01\\
  log $M_{\rm mol}/M_{\odot} <  6.5$    &  57.0 & 7.1 &  48.0 &  2.6 & 0.3 &  2.0 & 5.7 & 0.4 & 5.4 &  $-$1.39$\pm$0.10 &   3.02$\pm$0.59
  \enddata

  \tablecomments{Basic statistical properties (radius, velocity
    dispersion, and molecular gas mass) for the clouds extracted by
    {\tt clumpfind}. We break up the cloud statistics into two groups,
    log~$M_{\rm mol}/M_{\odot} \geq 6.5$ and log~$M_{\rm
      mol}/M_{\odot} < 6.5$, to show that there are significant
    differences between the general properties for these two
    populations. For $r$ and $\sigma_v$, we list the mean, the error
    on the mean, and the median as calculated using the Kaplan-Meier
    product-limit estimator to account for the upper limits.  The last
    two columns show the results from the ordinary least squares fits
    to the differential mass function for the two populations (Figure
    \ref{fig.massfunc}) in log-log space, where log d$N$/d$M_{\rm
      mol}$ = $\beta + \alpha$log~$M_{\rm
      mol}$.\label{table.properties}}
\end{deluxetable*}


\subsection{Molecular Gas and Cloud Kinematics}\label{section.kinematics}
\begin{figure*}
\hspace{-1cm}
\includegraphics[scale=.95,angle=0]{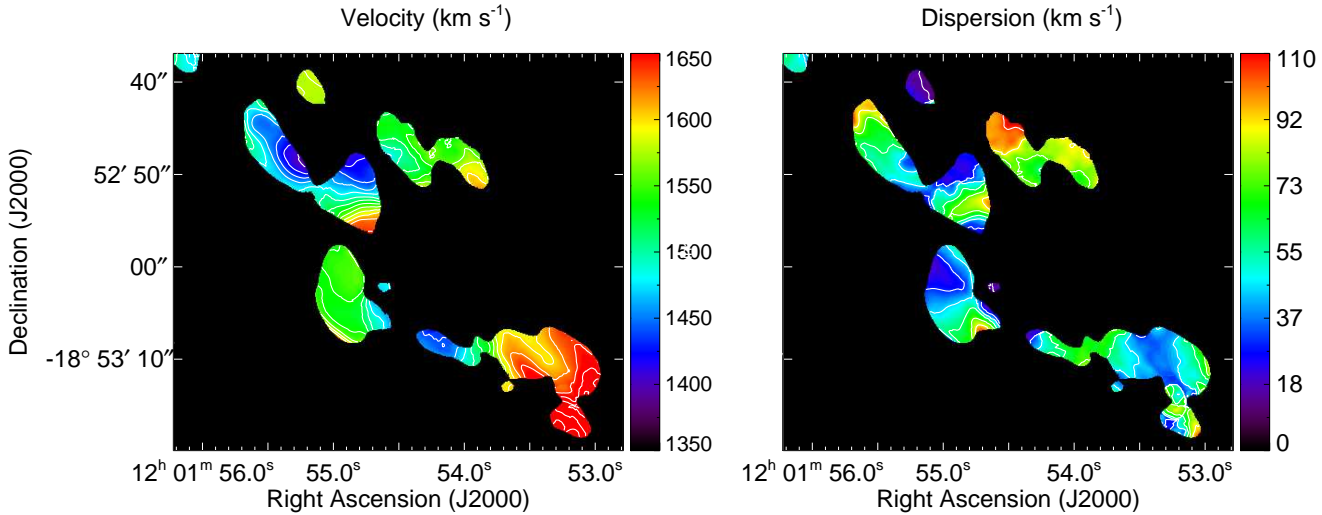}
\caption{Intensity-weighted velocity (left) and velocity dispersion
  (right) maps from our combined SMA$+$PdBI CO(2--1) data, masked at
  the 3\,$\sigma$ level of CO intensity. White contours demarcate
  velocities and velocity dispersions in steps of 20 and
  15\,km\,s$^{-1}$, respectively.\label{fig.mom1mom2}}
\end{figure*}

We show the intensity-weighted velocity and velocity dispersion map of
our combined SMA$+$PdBI CO(2--1) data in Figure
\ref{fig.mom1mom2}. Consistent with previous interferometric studies
of molecular gas in the Antennae \citep{gao01,wilson03,ueda11}, we
find that the kinematics of the molecular gas in the overlap region
are extremely complicated. Several of the CO(2--1) intensity peaks,
especially NGC~4039 in the lower right corner, show fairly obvious
velocity gradients. Intensity-weighted velocity dispersions are large
--- often $>$50\,km\,s$^{-1}$ --- most likely due to multiple clouds
along the line of sight, which we will discuss later. This does not
contradict our conclusion that we can resolve individual clouds
because {\tt clumpfind} differentiates by velocity as well as
spatially.

\begin{figure*}
\hspace{.5cm}
\includegraphics[scale=.8,angle=0]{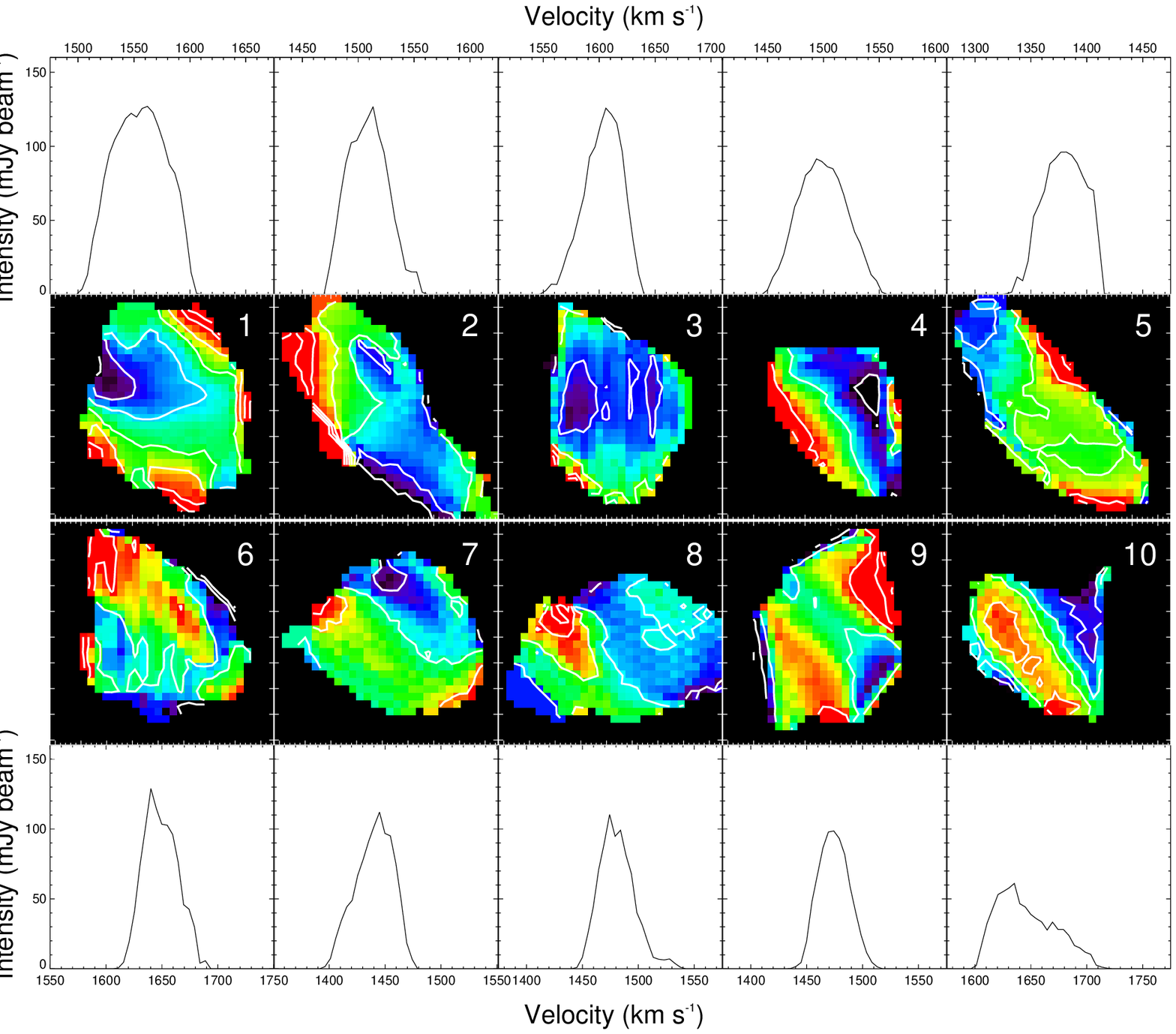}
\caption{Intensity-weighted velocity maps (middle panels) of the 10
  most massive clouds extracted by {\tt clumpfind} (labeled in Figure
  \ref{fig.big_clumps}), showing velocity gradients indicative of
  shear or rotation in many of the clouds. Clouds are numbered
  according to decreasing CO intensity. The angular size of each panel
  is the same (8.3\asec$\times$8.3\asec), but note that {\it the
    color-scale is different for each panel}, chosen to best show the
  velocity gradient. The white contours demarcate velocities in steps
  of 10\,km\,s$^{-1}$ for all panels. The CO spectrum of each cloud is
  plotted in the top/bottom panels, with channel widths of
  4.9\,km\,s$^{-1}$. \label{fig.clumps_velo}}
\end{figure*}

The velocity map gives the {\it average} velocity at any given
position, weighted by the CO intensity, so velocity information could
be blurred out if there are multiple clouds along the line of
sight. To preserve some velocity information, we made individual first
moment maps for the 50 brightest clouds extracted by {\tt clumpfind},
using the masks provided by the algorithm. Figure
\ref{fig.clumps_velo} shows the first moment velocity maps and spectra
for the 10 most massive clouds found by {\tt clumpfind} in our
combined SMA$+$PdBI data, with their spatial locations labeled in
Figure \ref{fig.big_clumps}. We find that about half of the clouds
show a velocity gradient indicating shear or rotation. Cloud 10 appears
to have multiple velocity components. However, because the velocity
peaks are separated by less than 2\,$\sigma_{\rm chan}$, {\tt
  clumpfind} assigns the emission to a single cloud.

\subsection{Evidence for Gravitationally Bound Clouds}\label{section.bound}

\begin{figure}
\hspace{-.5cm}
\includegraphics[scale=.6,angle=0]{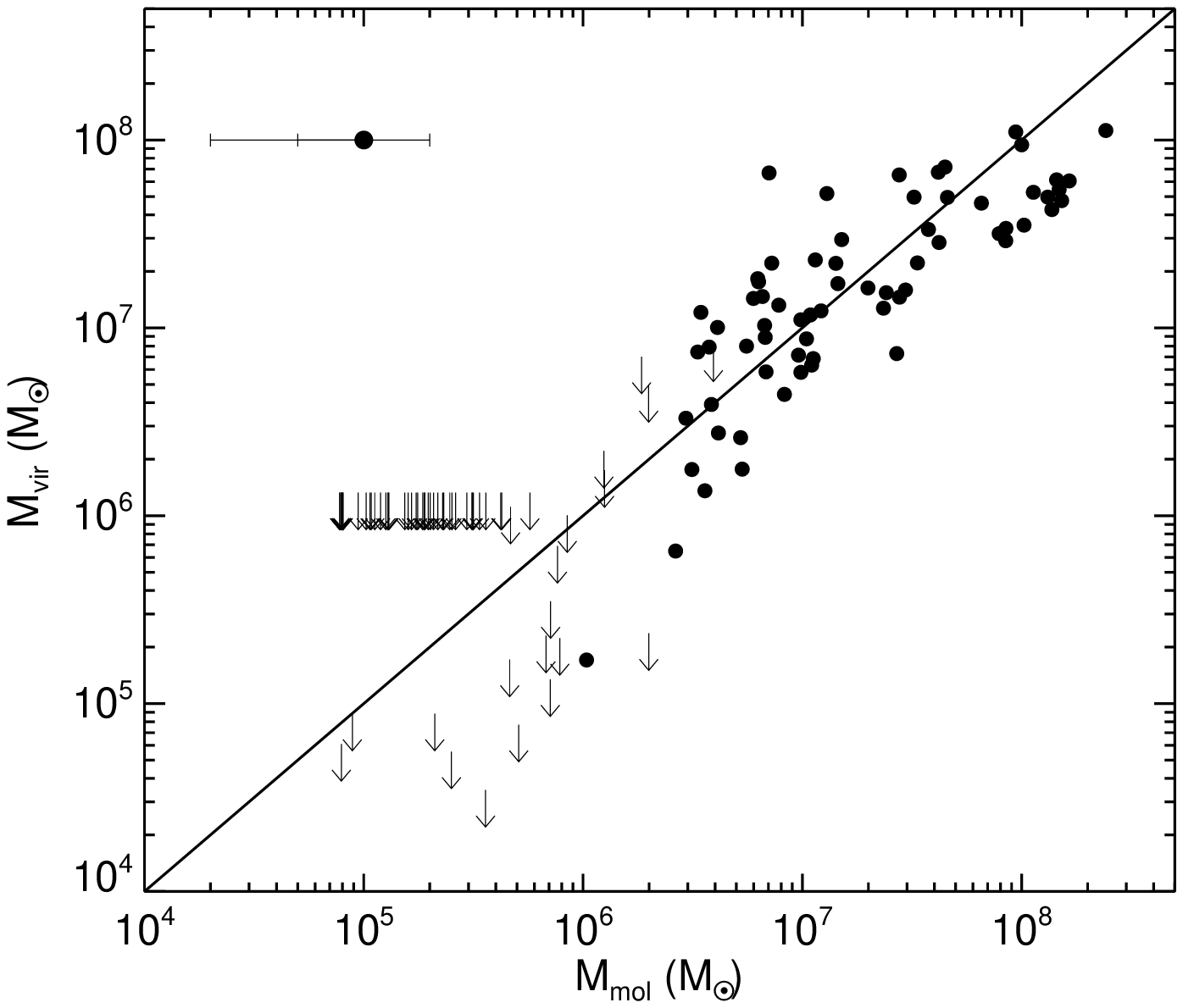}
\caption{Molecular gas mass estimated from the CO(2--1) intensity
  versus the virial mass ($M_{\rm vir} =$ 5$r\sigma_{v}^2/\alpha_g G$)
  estimated from the radius $r$ and velocity dispersion $\sigma_v$
  found by {\tt clumpfind}. We assume a geometrical factor $\alpha_g$
  of 5/3, which corresponds to a density profile scaling as
  $r^{-2}$. Arrows indicate upper limits in radius and/or velocity. We
  show in the upper left corner the distance each point would move to
  the left if $X_{\rm CO}$ was 2 and 5 times smaller, and to the right
  if $X_{\rm CO}$ was 2 times larger than our assumed value of $X_{\rm
    CO} =$ 3$\times$10$^{20}$ (K\,km\,s$^{-1}$)$^{-1}$. Solid line
  demarcates the one-to-one correspondence between the molecular gas
  and virial mass.  Most of our points fall on or below the line,
  suggesting that the clouds are gravitationally bound and some may be
  collapsing. \label{fig.virial}}
\end{figure}

We estimate the virial mass for each cloud from the resolution-corrected 
radius and velocity dispersion.  The virial mass is given by

\begin{equation}
M_{\rm vir} = \frac{5\, r\, \sigma_v^2}{\alpha_g \,G}, 
\end{equation}

\noindent where $\alpha_g$ is a geometric factor that depends on the
density profile.  For a self-gravitating cloud in virial equilibrium,
we expect the density profile to scale as $r^{-2}$
\citep{bodenheimer68, shu77, keto10} and $\alpha_g$ = 5/3. We plot the
relationship between molecular gas mass and virial mass in Figure
\ref{fig.virial}.  The detections alone give $\langle M_{\rm
  vir}/M_{\rm mol} \rangle = 1.2$, but this ratio drops to $\langle
M_{\rm vir}/M_{\rm mol} \rangle = 1.03$ if we properly take the upper
limits into account.

As mentioned in \S~\ref{section.massflux}, there is some debate over
whether $X_{\rm CO}$ is much smaller in the Antennae
\citep{zhu03,wilson03,schulz07}. We show in the upper left corner of
Figure \ref{fig.virial} the distance each point would move to the left
if $X_{\rm CO}$ was 2 and 5 times smaller, and to the right if $X_{\rm
  CO}$ was 2 times larger than our assumed value of $X_{\rm
  CO}~=~3\times$10$^{20}$ (K\,km\,s$^{-1}$)$^{-1}$. Considering the
many uncertainties inherent in the determination of either $M_{\rm
  vir}$ or $M_{\rm mol}$, the data are entirely consistent with
$M_{\rm vir}/M_{\rm mol} \approx 1$.  In other words, {\it the clouds
  are gravitationally bound structures}.  This is a key point.  It
implies that the extreme properties of the largest clouds --- their
large sizes, internal velocities, and masses --- are unlikely to be
artifacts of spatial or velocity blending.  If they were, then it
would be a coincidence that the radii and velocity dispersions of the
clouds would conspire to have just the right combination to produce
$M_{\rm vir} \approx M_{\rm mol}$.

If we run {\tt clumpfind} with lower thresholds, we still find a
one-to-one correspondence between molecular gas and virial mass, but
some of the low-mass clouds fall quite a bit below this
relation. These clouds typically have $\sigma_{\rm v}~<
2$~km\,s$^{-1}$ and upper limits on radius. There are only one or two
clouds found using the 3$\sigma_{\rm chan}$ threshold with the same
behavior, which reinforces our earlier conclusion that some of the
additional low-mass clouds extracted using the 2 and 2.5$\sigma_{\rm
  chan}$ detection thresholds may not be significant above the noise.

\subsection{Cloud Scaling Relations}\label{section.scaling}

\citet{larson81} described the relationship between the size,
linewidth (velocity dispersion), mass, and density of molecular
clouds.  Here we examine whether our clouds fall on these scaling
relations and how they compare to typical GMCs from non-starburst
galaxies.

\begin{figure}
\hspace{-.5cm}
\includegraphics[scale=.6,angle=0]{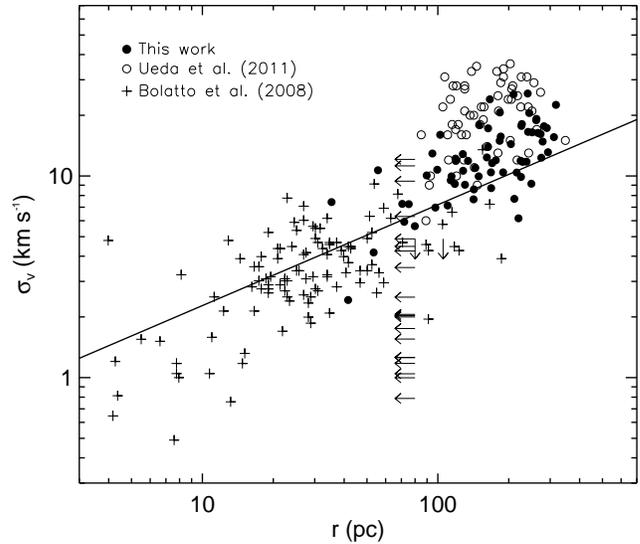}
\caption{Relationship between cloud radius ($r$) and velocity
  dispersion ($\sigma_v$). Clouds that were unresolved spatially or in
  velocity are assigned upper limits of 80\,pc (left-pointing arrow)
  and 4.9\,km\,s$^{-1}$ (downward arrow), respectively. Filled circles
  represent our data. Crosses represent GMCs from nearby galaxies
  (NGC~205, NGC 1569, NGC~3077, NGC~4214, NGC~4449, NGC~4605, IC~10,
  SMC, LMC, M33, M31) from \citet{bolatto08}. Open circles represent
  clouds from the Antennae found by \citet{ueda11} in CO(3--2). The
  solid line demarcates the typical Milky Way GMC relationship
  $\sigma_v \approx 0.72\,r^{0.5}$\,km\,s$^{-1}$
  \citep{solomon87}. \label{fig.sizeline}}
\end{figure}

We plot the relationship between radius ($r$) and velocity dispersion
($\sigma_v$) for our clouds in Figure
\ref{fig.sizeline}. \citet{solomon87} found a relationship of the form
$\sigma_v\,\propto\,r^{0.5}$ for Milky Way GMCs. This correlation has
been shown to hold even down to very small scales within individual
GMCs \citep{heyer09}. We find that most of our clouds fall on the
$r$--$\sigma_v$ relation, extending it to larger sizes and higher
velocity dispersions. However, the largest clouds in our study appear
to have systematically higher velocity dispersions than predicted for
Milky Way GMCs.  The clouds in the Antennae with $r$ \gax\ 100\,pc
have velocity dispersions higher by a factor of $\sim 2$ compared to
Galactic GMCs.  Clouds found by \citet{ueda11} in their CO(3--2)
observations of the Antennae, which were also obtained with the SMA
and at a resolution comparable to ours, also lie above the Galactic
$r$--$\sigma_v$ relation.  In fact, the CO(3--2) dispersions are even
larger than those measured in CO(2--1).  For the 35 clouds in common
between our sample and that of \citeauthor{ueda11}, $\langle
\sigma_v(3-2)/\sigma_v(2-1) \rangle \approx 2$. Matching our CO(2--1)
clouds with the CO(1--0) clouds from \citet{wilson03} is difficult
given the difference in resolution (3\farcs2$\times$4\farcs9 compared
to 3\farcs3$\times$1\farcs5), but we note that most of the CO(1--0)
clouds have velocity dispersions $<$10\,km\,s$^{-1}$, significantly
smaller than the velocity dispersion observed in the CO(3--2) and
CO(2--1) clouds.

The large offset in the \citeauthor{ueda11} velocity dispersions are
especially curious. The typical size-linewidth relation in Galactic
\citep{larson81,solomon87} and extragalactic \citep{bolatto08} clouds
predicts that clouds found by higher density tracers such as CO(3--2)
should have smaller velocity dispersions if they trace smaller size
scales, precisely the opposite of what \citet{ueda11} observe.

There are several possible explanations for the offset in velocity
dispersion for the largest clouds. Assuming that the Milky Way GMCs
represented by the solid line in Figure \ref{fig.sizeline} are in
virial equilibrium ($M_{\rm mol}\,\approx\,M_{\rm vir}\,\propto\,r\,
\sigma^2_{v}$), the size-linewidth relation
($\sigma_v\,\propto\,r^{0.5}$) found by \citet{solomon87} implies that
$M\,\propto\,r^2$ --- that all the Milky Way clouds have similar gas
surface densities. For the GMCs in the \citeauthor{solomon87} sample,
this corresponds to a gas surface density $\Sigma$ =
170\,$M_{\odot}$\,pc$^{-2}$. If the Antennae clouds are in virial
equilibrium, as they appear to be (Figure \ref{fig.virial}), then one
possibility is that they have higher surface densities than Milky Way
GMCs. \citet{gao01} find that the overlap region has exceptionally
high HCN (a high gas density tracer, $\ge\,10^{4}$~cm$^{-3}$)
emission, equivalent to half of the total HCN emission in the
system. Furthermore, the overlap region is extremely bright in the
far-infrared (FIR; \citealt{klaas10}), which is known to be
well-correlated with HCN \citep{gao04}. This correlation has been
shown to hold for a wide range of size scales, from dense Galactic
cores to extreme, high-redshift starbursts \citep{gao04,wu05,gao07}.

\begin{figure}
\hspace{-.5cm}
\includegraphics[scale=.6,angle=0]{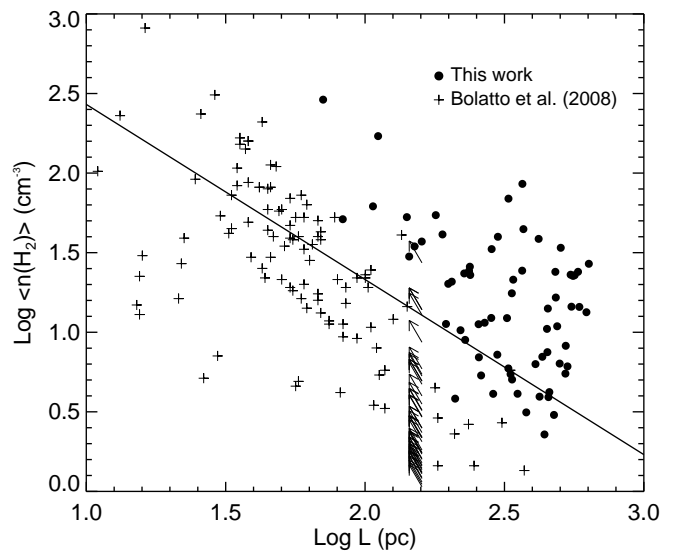}
\caption{Average H$_2$ number density, roughly estimated as $M_{\rm
    H_2}/(\frac{4}{3}\pi r^3)$, as a function of cloud diameter
  $L$. Crosses represent extragalactic GMCs from \citet{bolatto08}.
  The solid line denotes the relation $\langle n(\rm
  H_2)\rangle(cm^{-3}) = 3400\,$L$\,(pc)^{-1.1}$ found by
  \citet{larson81}.\label{fig.density}}
\end{figure}

We can roughly estimate the average H$_2$ number density in each cloud
from its CO brightness, assuming the mass is distributed uniformly
within a sphere. \citet{larson81} found that the average number
density ($\langle n(\rm H_2) \rangle$) varies inversely as the cloud
diameter $L$ [$\langle n({\rm H}_2)\rangle({\rm cm^3}) \propto L({\rm
    pc})^{-1.1}$], implying that the {\it column density} is
essentially independent of size. Figure \ref{fig.density} shows that
our clouds follow this relationship. At a given size, however, our
clouds appear to be offset toward higher number densities (by $\sim
0.3$ dex) than the molecular clouds from \citet{larson81}, as well as
GMCs from \citet{bolatto08}. This is consistent with the argument for
higher surface densities, and could also be the explanation for the
even larger offset of the \citet{ueda11} data.

Another possibility is the presence of elevated external pressure. In
the Galactic center, \citet{oka98,oka01} find GMCs that have extremely
large velocity dispersions --- approximately a factor of 5 larger at a
given radius than the clouds in the Galactic disk. \citet{oka98} argue
that the large velocity dispersions from turbulence keep the clouds in
equilibrium with high external pressures that arise from hot gas
and/or magnetic fields in this region. The overlap region of the
Antennae may also be experiencing higher pressures as the region of
contact between two gas-rich galaxies undergoing a major merger.  The
shocks produced under such violent conditions, especially in the
overlap region, could generate high interstellar pressures and high
turbulence. Higher $J$ transitions of CO are more sensitive to shocked
gas, and hence the higher velocity dispersions of the
\citeauthor{ueda11} clouds may be due to the increased sensitivity of
CO(3--2) to collisions and turbulence.

\citet{bolatto08} attribute the GMCs that have higher velocity
dispersions in their sample to blending of multiple clouds.  This,
too, is a distinct possibility in our data, given the complex geometry
and kinematics of the overlap region in the Antennae. Recent work has
shown that {\tt clumpfind} has difficulty accurately measuring cloud
parameters in crowded fields (e.g., \citealt{sheth08,pineda09}). A few
of our spectra in Figure \ref{fig.clumps_velo} show hints of multiple
peaks that could be interpreted as incomplete deblending by {\tt
  clumpfind}.  Nevertheless, as argued in \S~\ref{section.bound}, we
do {\it not}\ believe that blending seriously compromises our results,
although it is difficult to rule out this effect entirely.

At this point, it is unclear what is the primary cause for the larger
velocity dispersions --- higher gas densities, elevated external
pressure, collisions, turbulence, cloud blending, or a combination of
these factors. Observations at much higher angular resolution,
especially of optically thin CO species (C$^{17}$O, C$^{18}$O,
$^{13}$CO) and other high-density tracers (e.g., HCN, HCO$^+$) will
help distinguish between these different possibilities.

\subsection{The Mass Function}\label{section.massspec}

\begin{figure}
\hspace{-.5cm}
\includegraphics[scale=.6,angle=0]{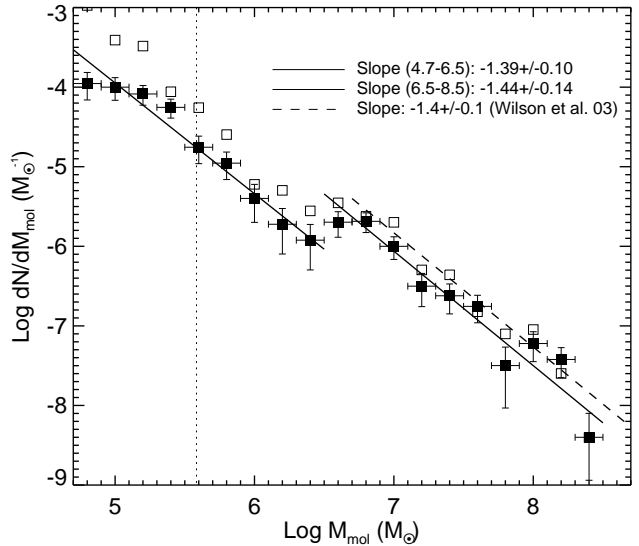}
\caption{Differential mass function for the clouds extracted by {\tt
    clumpfind} from our combined SMA$+$PdBI data cube. Filled black
  squares show the mass function for clouds found above
  3\,$\sigma_{\rm chan}$, and open squares show the mass function for
  clouds found above 2\,$\sigma_{\rm chan}$, showing that the break in
  the mass function is robust at both levels. Unsurprisingly, more
  clouds are found at the lower mass range using the 2\,$\sigma_{\rm
    chan}$ lower level, resulting in a steeper mass function for that
  half of the spectrum. Horizontal error bars indicate bin sizes;
  vertical error bars indicate $\sqrt(N)$ uncertainty in each bin. The
  vertical dotted line demarcates the 5\,$\sigma$ completeness limit
  of our data, 3.8$\times\,10^5\,M_{\odot}$. We fit two (solid) lines
  to the 3\,$\sigma_{\rm chan}$ mass function, one for bins with
  4.7$\leq$log~$M_{\rm mol}/M_{\odot}<$~6.5 and a second for bins
  with 6.5$\leq$log~$M_{\rm mol}/M_{\odot} \leq$~8.5. Both fits have
  slopes that are consistent with that found by \citet{wilson03}
  (dashed line).
\label{fig.massfunc}}
\end{figure}

We plot the differential mass function (d$N$/d$M_{\rm mol}$) of the
clouds found in the SMA$+$PdBI data in Figure \ref{fig.massfunc}, with
bins of log~$M_{\rm mol}/M_{\odot}$ = 0.2. The slope of the ordinary
least squares fit ($\alpha$) corresponds to a power law of the mass
function in linear space (d$N$/d$M_{\rm mol}$ $\propto$ $M_{\rm
  mol}^{\alpha}$). The improved sensitivity of our observations allows
us to extend the mass function down to lower masses than previous
work, down to a 5\,$\sigma$ completeness limit of
3.8$\times\,10^5\,M_{\odot}$\footnote{We estimate the 5$\sigma$
  completeness limit following the method described in
  \citet{wilson03}.}. By comparison, \citet{wilson03} reached a
completeness limit more than 10 times higher,
5$\times\,10^6\,M_{\odot}$ (6.7$\times\,10^6\,M_{\odot}$ if they
adopted a distance of 22\,Mpc).

One of the most remarkable features of the mass function is that there
is a distinct break around log~$M_{\rm mol}/M_{\odot}\approx 6.3-6.7$,
which coincides with the most massive Galactic clouds
\citep{solomon97}. This break appears to be robust with respect to the
detection threshold, showing up in both the 2 and 3\,$\sigma_{\rm
  chan}$ levels. Detailed examination of the ($u,v$) coverage and
amplitude versus ($u,v$) distance of our SMA and PdBI observations
does not reveal any gaps that could introduce a break in the mass
function. The break corresponds to the ``gap'' in the distribution of
masses shown in Figure \ref{fig.threepanel} and is the basis for our
motivation to divide the cloud population into two subsamples.

For the sake of concreteness, we chose the threshold to be log~$M_{\rm
  mol}/M_{\odot} = 6.5$. We perform two fits: one in the range
4.7~$\leq$log~$M_{\rm mol}/M_{\odot}$$<$~6.5, and a second in the
range 6.5~$\leq$log~$M_{\rm mol}/M_{\odot}$$\leq$~8.5. We find $\alpha
= -1.39\pm\,0.10$ for the first fit and $\alpha = -1.44\pm\,0.14$ for
the second fit. Figures \ref{fig.threepanel}c and \ref{fig.massfunc}
indicate that the mass distributions of the two populations may be
log-normal, dropping off at the lower-mass end of each range. If we
exclude the two end points (log~$M_{\rm mol}/M_{\odot} = 4.8$ and 6.5)
where the two distributions appears to turn over, the slopes for both
the low- and high-mass ranges are then steeper, with $\alpha =
-1.52\pm\,0.08$ and $\alpha = -1.52\pm\,0.14$, respectively. These
slopes are all statistically consistent with $\alpha = -1.4\pm\,0.1$
found by \citet{wilson03}.

Note that if we use a distance of 19\,Mpc to be completely consistent
with \citet{wilson03}, the break remains apparent at log~$M_{\rm
  mol}/M_{\odot}\sim$~6.3 and the respective fits have slopes $\alpha
= -1.64\pm\,0.16$ and $\alpha = -1.40\pm\,0.15$. Additionally, we find
that if we run {\tt clumpfind} with the lower detection thresholds,
the slopes of the mass function at the low-mass end increases slightly
(from $-$1.39 to $-$1.68), while the slope of the high-mass end
remains relatively constant within the errors. We find that the break
between the low- and high-mass differential mass function also remains
constant with different detection thresholds; it remains log~$M_{\rm
  mol}/M_{\odot}\approx 6.3-6.7$ for all three thresholds that we
tested.

The slopes of our mass function are well-matched to the slope of the
mass function for young clusters in the Antennae, $\alpha \approx
-2.0$ (\citealt{zhang99,whitmore10}). This is consistent with the good
agreement between the GMC mass spectrum and the upper range of the SSC
mass spectrum in M82 \citep{keto05}, suggesting that individual star
clusters may be closely related to individual clouds.

Our fit over the mass range 6.5~$\leq$log~$M_{\rm
  mol}/M_{\odot}$$<$~8.5 suggests that some of the largest super-giant
molecular complexes found by \citeauthor{wilson03} break up into
smaller clouds in our high-resolution observations. Importantly, many
of the large clouds {\it remain}\ large despite the improved spatial
resolution of our observations, which is sufficient to resolve the
larger structures even at our highest resolution. The largest clouds
in our data are much larger than GMCs in our Galaxy and typical
star-forming galaxies (1--100\,pc; \citealt{solomon87,bolatto08}).

Returning to Figure \ref{fig.threepanel} and Table 3, we see that the
two populations below and above the break have statistically different
properties. While the clouds with 6.5~$\leq$log~$M_{\rm
  mol}/M_{\odot}$$<$~8.5 tend to be large ($r$ \gax\ 100\,pc) and have
high velocity dispersions ($\sigma_v$ \gax\ 10 km\,s$^{-1}$), the less
massive clouds tend to be largely unresolved in both radius ($r$
\lax\ 80\,pc) and velocity dispersion ($\sigma_v$ \lax\ 4.9
km\,s$^{-1}$). Also intriguing is the {\it spatial distribution}\ of
the clouds in Figure \ref{fig.big_clumps} --- the more massive clouds
are associated with peaks of intense star formation
(\S~\ref{section.sf}), while the population of smaller, less massive
GMCs resides in more quiescent areas far from the peaks.

In summary, these results suggest a bimodal distribution of giant
molecular clouds/complexes in the Antennae: a regular, quiescent
population that is similar to typical GMCs in nearby galaxies and the
Milky Way, and a second population that appears much larger, more
turbulent, and unusually massive.  The exceptional properties of the
largest clouds appear connected to the extreme conditions associated
with the interaction zone of the merging galaxies.  The dividing line
between the two populations occurs near $M_{\rm mol} \approx
3\times10^6\, M_\odot$, which, interestingly, coincides roughly with
the upper end of molecular cloud masses seen in the Milky Way
\citep{solomon97}.

\section{Comparison with Star Formation Tracers}\label{section.sf}
We explore the relationship between dense molecular gas and star
formation by comparing the CO emission with various tracers of recent
and on-going star formation.

\subsection{Super Star Clusters}

\begin{figure*}
\vspace{-3cm}
\hspace{-1cm}
\includegraphics[scale=1,angle=0]{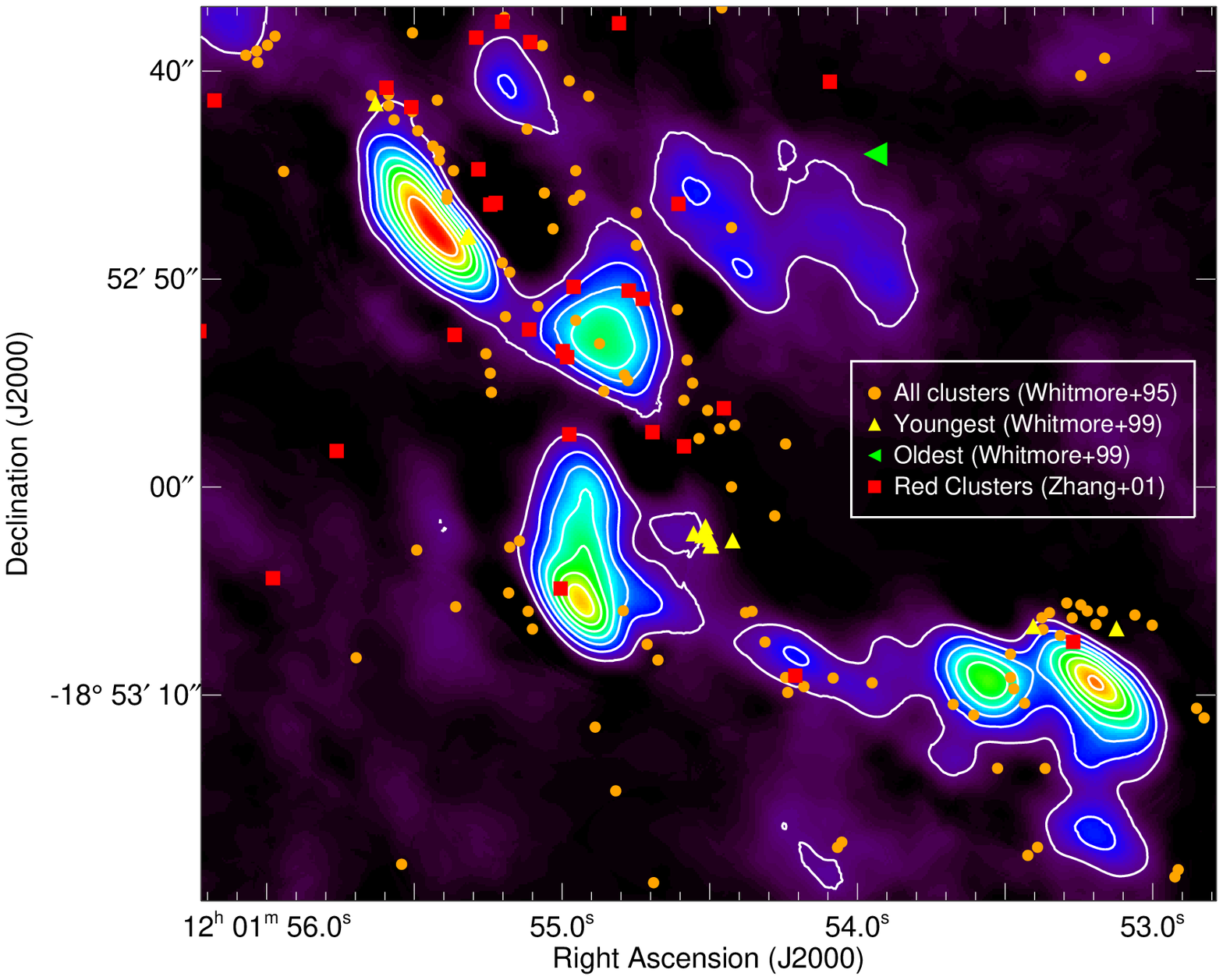}
\vspace{-3cm}
\caption{Velocity-integrated CO(2--1) emission from the overlap region
  of the Antennae, with the same contours as Figure
  \ref{fig.big_clumps}. Symbols mark the locations of SSCs of
  different ages identified by \citet{whitmore95} and
  \citet{whitmore99}, as well as red clusters from \citet{zhang01},
  both of which appear to preferentially surround the CO peaks. Note
  that the red clusters from \citet{zhang01} are candidates for the
  youngest clusters ($\sim$5\,Myr), as the reddening is thought to
  come from dust obscuration. \label{fig.big_clusters}}
\end{figure*}

As discussed in the Introduction, SSCs are extremely young ($\lesssim$
few -- 50 Myr), representing very recent star formation, and are quite
prominent in the Antennae. Figure \ref{fig.big_clusters} shows again
the velocity-integrated CO(2--1) emission, this time with the
locations of various SSCs overplotted. These clusters are identified
as compact sources in {\it HST}\ optical images
\citep{whitmore95,whitmore99,zhang01}. All our CO clouds have SSCs
nearby. However, the SSCs are identified in the optical and are not
good tracers of embedded or current star formation.

\subsection{Current Star Formation}\label{section.tracers}

We also consider the relationship between current star formation and
the distribution and intensity of molecular gas, velocity gradients,
and other signatures of shocks and compression in our CO data that may
provide insight into the nature of SSC formation. We focus on three
star formation tracers with the highest spatial resolution available
--- H$\alpha$ emission from the recombination of gas ionized by young,
massive O and B stars, 8 $\mu$m emission from polycyclic aromatic
hydrocarbons (PAHs) heated by star formation, and 4\,cm radio
continuum from either H~II regions or young supernova remnants. The
H$\alpha$ data were taken with the F658N filter on the {\it HST} Wide
Field Planetary Camera 2 (WFPC2) by \citet{whitmore99}, the PAH maps
were acquired using the {\it Spitzer} Infrared Array Camera (IRAC) by
\citet{wang04}, and the 4\,cm continuum map was taken by
\citet{neff00} with the Very Large Array (VLA).

\citet{wang04} estimate that $\sim$5\% of the 8 $\mu$m emission comes
from diffuse gas heated by older stars, so we subtract this component
from the 8\,$\mu$m image following \citet{pahre04ch1}, by scaling the
3.6 and 4.5 $\mu$m channels to match the colors of M0 III
stars. Comparison between 4\,cm continuum emission and 8 $\mu$m PAH
emission shows very good spatial correlation between the two tracers
(Figure \ref{fig.sftracers}), so we elect to show only the 8 $\mu$m
PAH emission in the next figure. Figure \ref{fig.sftracers} also notes
the location of peaks in the FIR emission at 70$\mu$m (from the {\it
  Herschel} Photodetector Array Camera \& Spectrometer -- PACS;
\citealt{klaas10}), showing good agreement with the 8$\mu$m PAH and
4\,cm continuum emission within the 5\farcs5 resolution of the FIR
image.

\begin{figure}
\vspace{-1cm}
\includegraphics[scale=.45,angle=0]{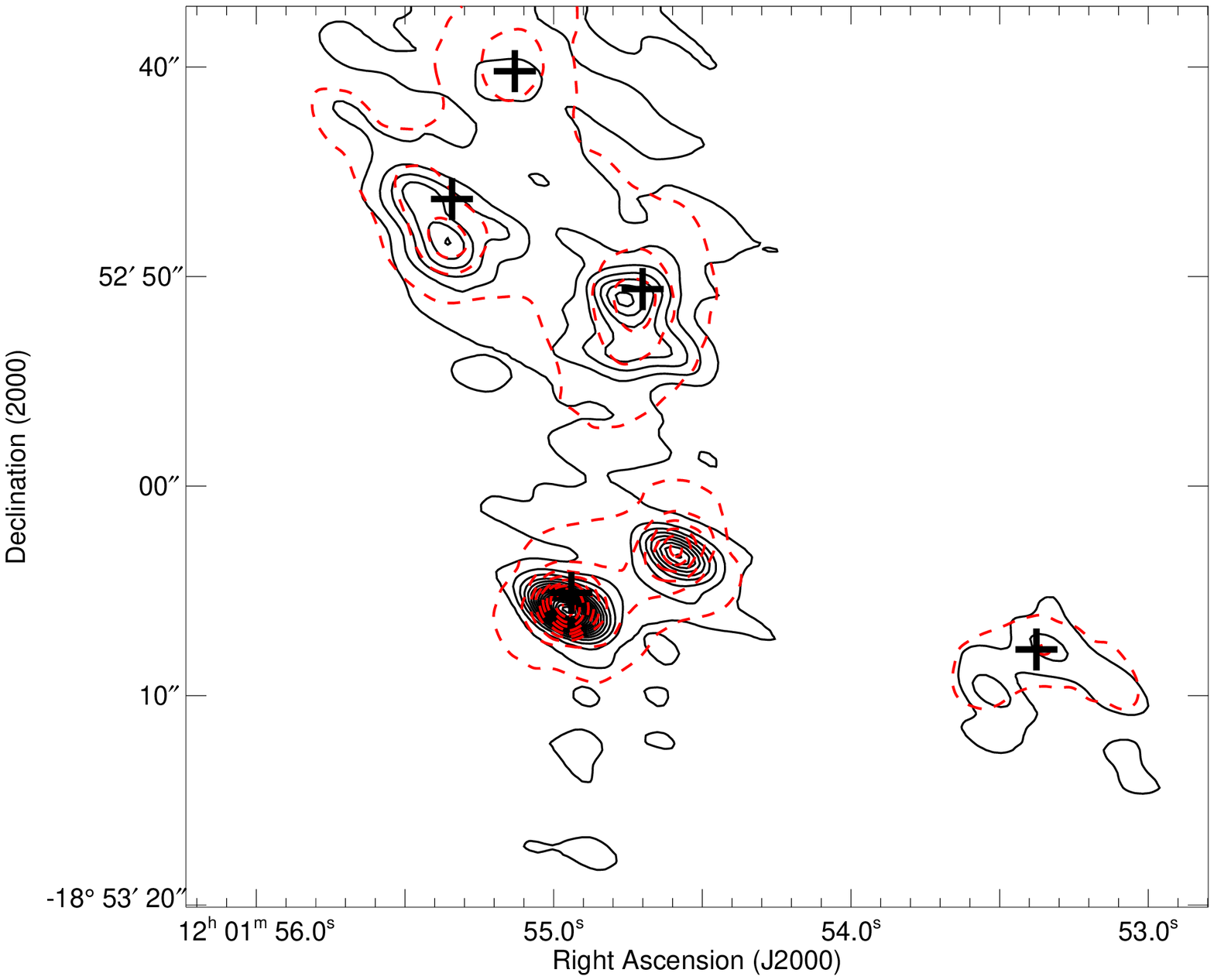}
\vspace{-.7cm}
\caption{Contours of 4\,cm continuum (black; Neff \& Ulvestad 2000)
  and 8$\mu$m PAH emission (red; Wang et al. 2004) in the overlap
  region of the Antennae, showing very good agreement between the two.
  Contour levels range from 5\%--95\% of peak intensity in steps of
  5\% for the 4\,cm continuum emission and 10\%--90\% of peak
  intensity in steps of 10\% for the 8\,$\mu$m PAH emission. FIR
  emission peaks from {\it Herschel} PACS 70$\mu$m imaging are marked
  with crosses, consistent with the 8$\mu$m PAH and continuum peaks
  within its 5\farcs5 resolution limit. Note that the absence of a FIR
  counterpart for the central 8$\mu$m PAH/4\,cm continuum peak is due
  to the coarser resolution of the 70$\mu$m imaging blending the two
  peaks together. \label{fig.sftracers}}
\end{figure}

\begin{figure*}
\includegraphics[scale=.9,angle=0]{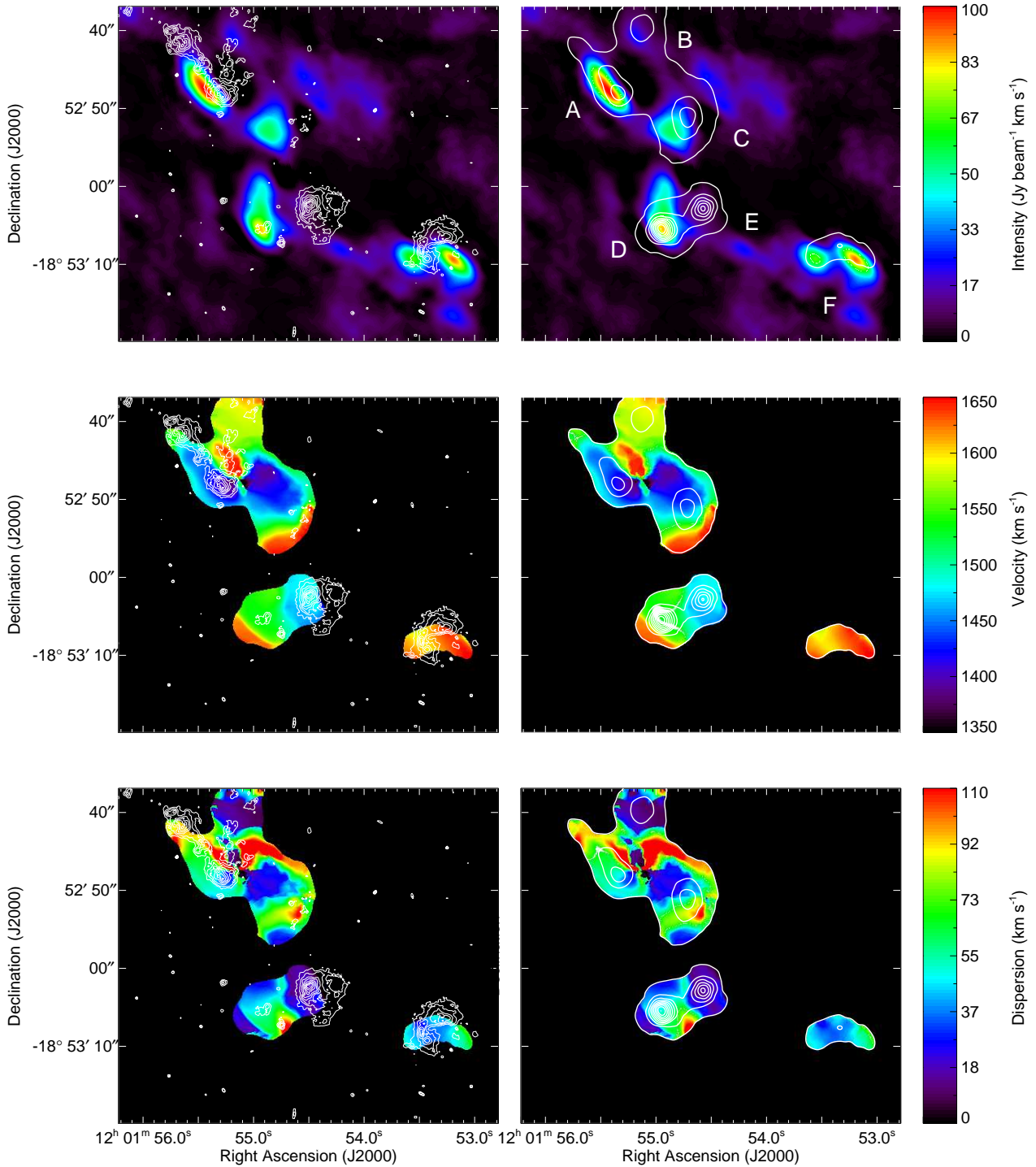}
\vspace{-3.cm}
\caption{\footnotesize Color-scale images of the velocity-integrated
  emission (zeroth moment), intensity-weighted velocity (first
  moment), and intensity-weighted velocity dispersion (second moment)
  from our combined SMA$+$PdBI CO(2--1) data from top to bottom,
  respectively. The first and second moments are masked below 10\% of
  peak PAH intensity to suppress the noise in regions of weak CO
  emission. Color bars on the right indicate the range displayed in
  each panel; left and right panels are identical. The left panels are
  overlaid with contours of H$\alpha$ emission \citep{whitmore99}, and
  the right panels are overlaid with contours of 8\,$\mu$m PAH
  emission \citep{wang04}, showing the distribution of star
  formation. Contour levels range from 0.02--1.57 counts\,s$^{-1}$ in
  equal steps of 0.24 dex for the H$\alpha$ emission and 10\%--90\% of
  peak intensity in steps of 10\% for the 8\,$\mu$m PAH emission. The
  star-forming peaks discussed in the text and Figure \ref{fig.big_pv}
  are labeled A--F in the top right panel.\label{fig.big_moments}}
\end{figure*}

\begin{figure*}
\includegraphics[scale=.8,angle=0]{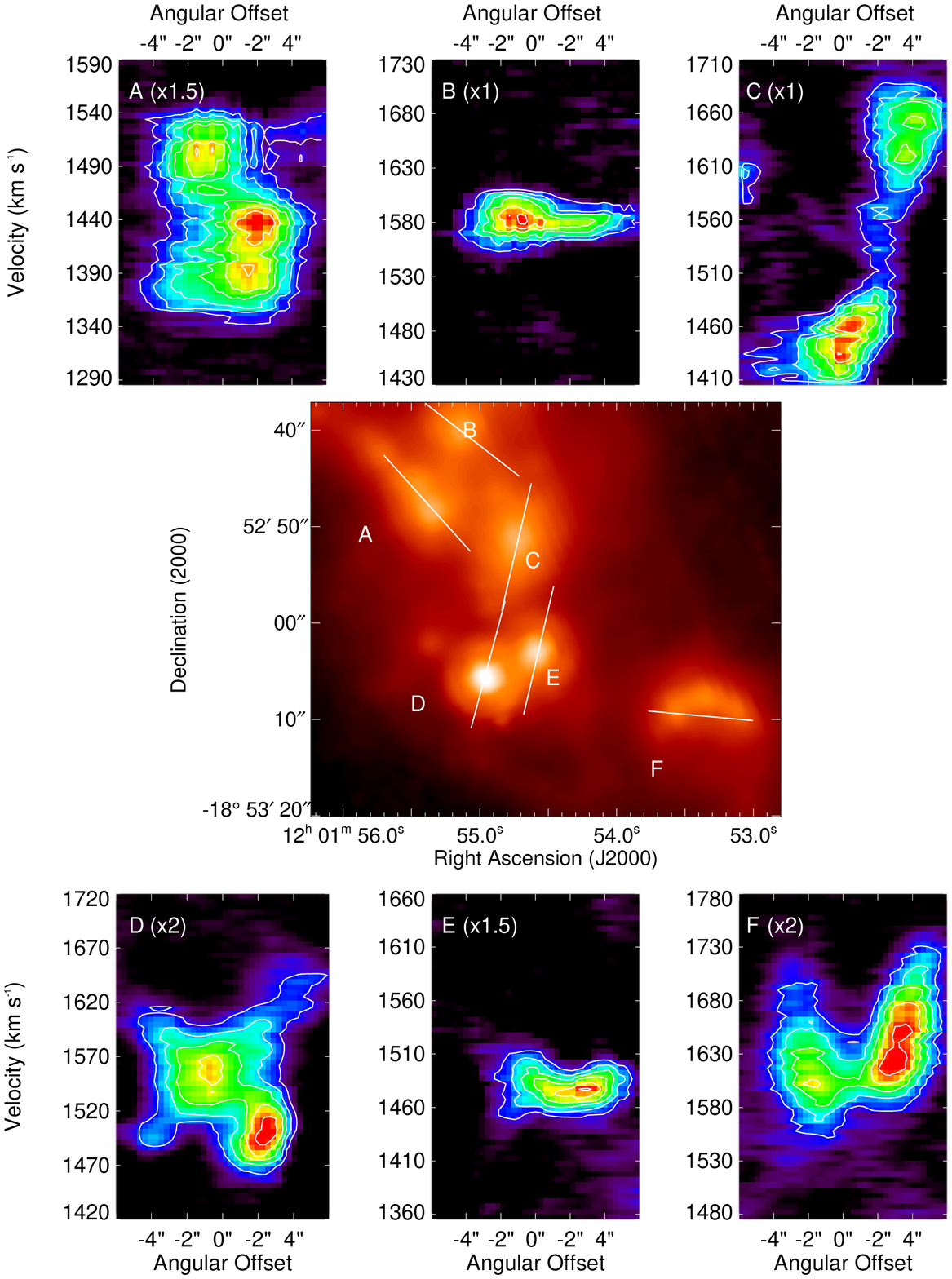}
\vspace{-.25cm}
\caption{\footnotesize Position-velocity diagrams of CO(2--1) emission
  (upper and lower panels) from our combined SMA$+$PdBI data, with
  cuts made at the star-forming peaks as traced by 8\,$\mu$m PAH
  emission (center panel in heat scale; the cuts are demarcated by
  white lines and labeled A--F to match Figure
  \ref{fig.big_moments}). Cuts are chosen to show the largest velocity
  gradient in the CO emission. The top and bottom of each slice
  corresponds to the left and right side of each panel,
  respectively. Contour levels are 20, 35, 50, 65, 80, and 95\% of the
  peak intensity of the cut with the weakest emission (Panel C,
  0.47\,Jy\,beam$^{-1}$), multiplied by a factor noted in each panel
  to better display the cuts with stronger emission. Color scale of
  each image is also adjusted accordingly. The position-velocity
  diagrams show that the structure of the molecular gas is often
  extremely complex at the star-forming peaks, showing multiple peaks
  and velocities spanning up to 300 km\,s$^{-1}$, suggesting that this
  is a very dynamically active region.
 \label{fig.big_pv}}
\end{figure*}

Figure \ref{fig.big_moments} shows the velocity-integrated emission
(zeroth moment; top), the intensity-weighted velocity (first moment;
middle), and the intensity-weighted velocity dispersion (second
moment; bottom) of our combined SMA$+$PdBI CO(2--1) data. Each panel
is overlaid with contours of H$\alpha$ (left) or 8 $\mu$m PAH (right)
emission to show the distribution of star formation.

As discussed in \S \ref{section.clumps}, there appear to be multiple
overlapping clouds along the line of sight in the CO-bright regions of
our data. This means that in a straight first moment map, much of the
velocity information is lost due to averaging. Another way to view the
velocity distribution of the molecular gas would be to consider
position-velocity slices of emission at each of the star-forming peaks
(Figure \ref{fig.big_pv}). We choose the position-velocity slices to
show the largest velocity gradients in the CO emission.

We discuss the prominent star-forming peaks in the 8 $\mu$m map, labeled
A--F in the top right panel of Figure \ref{fig.big_moments}, and
consider the molecular gas distribution and kinematics at these peaks.

{\it Peak A:} Star formation in this region is seen both in H$\alpha$
and PAH, both of which appear to be slightly offset from the peak CO
emission. The higher resolution H$\alpha$ image shows that the star
formation breaks up into two or three peaks, one of which is almost
completely offset from the CO. \citet{brandl09} estimate a SFR of 0.74
$M_{\odot}$\,yr$^{-1}$ for this whole region based on its infrared
luminosity. This is also one of the brighter regions in terms of
integrated CO intensity. The position-velocity diagram in Figure
\ref{fig.big_pv} shows two or three clouds spatially separated by
$\sim$3$^{\prime\prime}$ ($\sim$321\,pc). The spread in gas velocity
is very large, ranging from 1340 to 1540\,km\,s$^{-1}$, but there is
only the slightest hint of a velocity gradient for the upper left CO
cloud. Peak A is associated with water maser emission
\citep{brogan10}, indicating very young star formation.

{\it Peak B:} One of the fainter star-forming peaks based on the PAH
and 4cm emission, this region barely shows up in the H$\alpha$
imaging. The integrated CO emission is also one of the weakest,
although there may be two clouds at around the same velocity,
spatially separated by $\sim$3\farcs5 ($\sim$375\,pc).

{\it Peak C:} A region where the H$\alpha$ emission may be mostly
obscured, the 8 $\mu$m PAH peak is slightly offset from the CO
peak. This is one of the most highly obscured areas in the overlap
region, with $A_V = 10.3-11.8$ mag \citep{snijders07,mengel05}. The
SFR in this region is 0.66 $M_{\odot}$\,yr$^{-1}$
\citep{brandl09}. The velocity dispersion of the CO is extremely
large, $>$300\,km\,s$^{-1}$, with at least two clouds with large
velocity gradients, seen in both the middle panel of Figure
\ref{fig.big_moments} and the upper right panel of Figure
\ref{fig.big_pv}. Near-infrared spectroscopic imaging by
\citet{herrera11} also find extremely broad line widths in the warm
H$_2$ emission, which they attribute to shocks driven by gas dynamics
on large scales. A compact, warm H$_2$ source observed in this region
is thought to be associated with a massive cloud on its way to
becoming a SSC \citep{herrera11}.

{\it Peak D:} This is one of the brightest peaks in 8 $\mu$m PAH and
in CO(2--1). The large amount of dust and gas present explains why
this peak is almost completely obscured in the H$\alpha$
image. \citet{brandl09} estimate the SFR for this region to be large,
almost 2 $M_{\odot}$\,yr$^{-1}$. This region also contains a water
maser \citep{brogan10}.  Multi-wavelength spectral energy
distributions suggest that some of the youngest clusters in the
Antennae reside here \citep{zhang10}. The position-velocity diagram
shows at least two clouds\footnote{Note that the contours of CO
  emission are scaled from the faintest cloud, so the lowest contour
  here is 2$\times$ brighter than the lowest contour of peak B.}  with
large velocity gradients, possibly in the process of merging. The
molecular gas spans a large range in velocity,
$\sim$200\,km\,s$^{-1}$, with $\sim$80\,km\,s$^{-1}$ separation
between the two clouds.

{\it Peak E:} This peak has the brightest H$\alpha$ and second
brightest 8 $\mu$m PAH emission, but it contains surprisingly small
amounts of CO. \citet{brandl09} estimate a SFR of 1.8
$M_{\odot}$\,yr$^{-1}$ for this region from its infrared
luminosity. Figure \ref{fig.big_clusters} shows that the youngest SSCs
found by \citet{whitmore99} are clustered around this region. This
suggests that peak E is a site of very recent and vigorous star
formation, but given the lack of molecular gas present, star formation
will not be able to continue at the same rate for much longer. The
distribution of the remaining molecular gas is also interesting; the
curved shape in the position-velocity diagram is suggestive of gas
infalling toward the center.

{\it Peak F (NGC 4039):} This region contains the nucleus of NGC~4039,
one of the galaxies in the merging Antennae pair. Associated with one
of the water masers found by \citet{brogan10}, the age of star
clusters in this area appears quite young, less than 6.3\,Myr
\citep{bastian09}. The SFR from the infrared luminosity, however,
appears much lower than in the other peaks in the overlap region
($\sim$0.33 $M_{\odot}$\,yr$^{-1}$; \citealt{brandl09}). Activity from
an active galactic nucleus, if any, is extremely weak, and the
infrared emission appears to be dominated by purely young star
formation \citep{brandl09}. The distribution of the H$\alpha$ emission
is quite different and offset from the 8 $\mu$m PAH emission. There
are at least two large clouds in the CO(2--1) emission in this region,
spanning a velocity range of $\sim$200\,km\,s$^{-1}$. One of the
clouds (right) show a fairly obvious velocity gradient, while the
other (left) does not. The peaks are separated spatially by
$\sim$4$^{\prime\prime}$ ($\sim$428\,pc).

\subsection{Implications}

Our comparison of the molecular gas distribution and kinematics with
different star formation tracers (8 $\mu$m PAH, 4\,cm continuum, and
H$\alpha$) shows that the molecular gas at star-forming peaks have
large velocity dispersions, typically spanning velocities in the range
of 200--300\,km\,s$^{-1}$. In addition, we find that the molecular gas
at all but two of the star-forming peaks (B and E) have large velocity
gradients.

\citet{keto05} find that the clouds with active star formation in the
starburst galaxy M82 are experiencing shock-driven compression, with
both lateral and line-of-sight velocity gradients (i.e., inverse P
Cygni profiles) within the individual clouds. These observations
suggest that individual massive star clusters may have formed from
individual GMCs crushed by a sudden galactic-scale increase in
external pressure. While we do not observe signatures of inverse P
Cygni profiles in our data (the presence of multiple clouds along the
line of sight may blur out such signatures), the large velocity
gradients we observe in some of the star-forming peaks may reflect
compressive shocks.

Recent work provides additional observational evidence supporting the
pressure-triggered formation scenario. As mentioned in the previous
section, detailed examination of near-infrared emission in Peak C
suggest that the warm H$_2$ emission is excited by large-scale shocks
\citep{herrera11}. Based on the ages of star-forming regions,
\citet{zhang10} propose two paths of shock-induced sequential star
formation, perhaps propagated by the collision of GMCs, across the
overlap region of the Antennae. Both paths start from the center of
the overlap region (around Peak D in Figure \ref{fig.big_moments}) and
move outwards to the north-west and south-east regions. This is
well-correlated with the region with the highest star formation
efficiency \citep{gao01} and FIR emission \citep{klaas10}.

Additionally, \citet{whitmore10} find evidence for shock-triggered
sequential cluster formation on a small scale, {\it within} the
individual star-forming regions, based on the spatial distribution of
clusters of different ages. One example of this sequential cluster
formation highlighted by \citet{whitmore10} involves Peak E in our
work, where an older cluster (10--50\,Myr) to the right may have
triggered the formation of much younger star clusters to the left
(Figure \ref{fig.big_clusters}). And finally, simulations of the
merger between NGC~4038 and NGC~4039 suggest that compressive tides
occur frequently and play an important role in the evolution of the
Antennae \citep{renaud09}.

All this clearly ties in with our results from
\S~\ref{section.clumps}. The largest clouds are associated with
regions of intense infrared emission, active star formation, and high
star formation efficiencies
\citep{evans97,mirabel98,gao01,brandl09,klaas10}, where the molecular
gas appear to have large velocity dispersions and gradients
(\S~\ref{section.tracers}). These clouds are different than typical
GMCs in nearby star-forming galaxies. While they appear to be
gravitationally bound, many of these clouds are larger, more massive,
and have higher velocity dispersions than expected from typical GMC
scaling relations. Given that they reside in such an extreme
environment, some combination of higher gas densities, elevated
external pressure, turbulence, and cloud collisions is most likely
responsible for the formation of these massive clouds, resulting in
the bimodal distribution we observe in the cloud mass spectrum.

\section{Summary}\label{section.summary}

In this paper, we present high-resolution SMA and PdBI CO(2--1)
imaging of the intensely star-forming overlap region in the Antennae
galaxies in order to study the properties of molecular gas that
appears to be forming SSCs.

Using {\tt clumpfind}, we resolve the CO(2--1) emission into a large
number of clouds. This extends the cloud mass function down to lower
masses than previous work by \citet{wilson03}, down to a 5$\sigma$
completeness limit of 3.8$\times10^5\,M_{\odot}$. A break in the cloud
mass function around log~$M_{\rm mol}/M_{\odot}\approx 6.5$ suggests
two distinct population of clouds. Fitting a broken power law to the
mass function, we find that dN/d$M_{\rm mol}~\propto~M_{\rm
  mol}^{-1.39}$ for the low-mass cloud population and dN/d$M_{\rm
  mol}~\propto~M_{\rm mol}^{-1.44}$ for the high-mass cloud
population, close to the $M^{-2}$ mass function found for SSCs in the
Antennae. We find that the two populations have statistically
different properties, and appear to populate different areas in the
overlap region.

The low-mass population of clouds (log~$M_{\rm mol}/M_{\odot} < 6.5$),
often below our spatial and velocity resolution limit, appears to have
properties similar to GMCs in typical star-forming galaxies. These
clouds tend to be distributed along the outskirts of the overlap
region, away from sites of intense star formation.

The high-mass cloud population (log~$M_{\rm mol}/M_{\odot} \geq 6.5$),
is associated with sites of peak star formation. These clouds appear
to be much larger in size (median $r~=$~170\,pc) and mass (median
log~$M_{\rm mol}/M_{\odot}~=$~7.2) compared to typical GMCs. While we
find that these clouds appear to be in virial equilibrium ($M_{\rm
  vir}~\sim~M_{\rm mol}$), they have higher velocity dispersions in
the size-linewidth relation than expected for typical GMCs.

Comparison between our CO data and star formation traced by 8 $\mu$m
PAH/4\,cm continuum/H$\alpha$ imaging show good correlation between
the distribution of molecular gas and current star
formation. Consistent with observational evidence of
pressure-triggered star formation in the region, position-velocity
slices of the CO data reveal large velocity gradients and dispersions
in the gas at the star-forming peaks, indicative of an active,
turbulent environment where compressive shocks may dominate. These
results suggest that factors such as elevated external pressure,
increased turbulence, and cloud collisions are most likely responsible
for the apparent high velocity dispersions and gas densities,
resulting in the formation of the high-mass clouds and complexes we
observe in the overlap region. These high-mass clouds and complexes
may, in turn, form the SSCs that are so prevalent in the Antennae.

While our observations have allowed us to probe the properties of the
molecular gas at much higher angular resolution and sensitivity than
previous work, we are still limited to size scales much larger than
that of typical star clusters ($<$~10\,pc). With the advent of the
full extended array with ALMA, sensitive and very high-resolution
($<$0\farcs1) imaging of the ISM in the Antennae will show whether the
large, massive clouds we observe will begin to break down into smaller
objects and steepen the cloud mass function.

\acknowledgements We thank the anonymous referee for his/her helpful
comments which improved this manuscript. We thank Zhao-Yu Li for help
in generating the three-color image in Figure \ref{fig.hst}. We are
grateful to Alberto Bolatto, Izaskun Jimenez-Serra, and Peter Teuben
for helpful discussions. The research of L.C.H. is supported by the
Carnegie Institution for Science. The Submillimeter Array is a joint
project between the Smithsonian Astrophysical Observatory and the
Academia Sinica Institute of Astronomy and Astrophysics and is funded
by the Smithsonian Institution and the Academia Sinica. This work is
based in part on archival data obtained with the Spitzer Space
Telescope, which is operated by the Jet Propulsion Laboratory,
California Institute of Technology under a contract with NASA.

\clearpage

\LongTables
\tablenum{2}
\begin{deluxetable*}{lllrrrr}
  \tabletypesize{\scriptsize}
  \tablewidth{0pt}
  \tablecaption{Cloud Catalog\label{table.clumps}}

  \tablehead{ 
    \colhead{} & 
    \colhead{$\alpha_{2000}$,$\delta_{2000}$} &
    \colhead{$v_{\rm rad}$} &
    \colhead{$r$} & 
    \colhead{$\sigma_v$} & 
    \colhead{$\log M_{\rm mol}$} &
    \colhead{$\log M_{\rm vir}$}
    \\ 
    \colhead{} & 
    \colhead{} &
    \colhead{(km\,s$^{-1}$)} &
    \colhead{(pc)} &
    \colhead{(km\,s$^{-1}$)} & 
    \colhead{($\log M_{\odot}$)} & 
    \colhead{($\log M_{\odot}$)}
  }

  \startdata
       1 &  12$^{\rm h}$01$^{\rm m}$53$^{\rm s}.$22,$-$18$^{\circ}$53$^{\prime}$09$\farcs$5 & 1615.7 &   251 &   16.5 &  8.2 &    7.7\\
       2 &  12$^{\rm h}$01$^{\rm m}$55$^{\rm s}.$50,$-$18$^{\circ}$52$^{\prime}$47$\farcs$9 & 1508.4 &   290 &   17.3 &  8.2 &    7.8\\
       3 &  12$^{\rm h}$01$^{\rm m}$54$^{\rm s}.$95,$-$18$^{\circ}$53$^{\prime}$05$\farcs$6 & 1493.8 &   183 &   20.6 &  8.2 &    7.7\\
       4 &  12$^{\rm h}$01$^{\rm m}$55$^{\rm s}.$38,$-$18$^{\circ}$52$^{\prime}$48$\farcs$5 & 1430.4 &   163 &   17.2 &  7.9 &    7.5\\
       5 &  12$^{\rm h}$01$^{\rm m}$53$^{\rm s}.$18,$-$18$^{\circ}$53$^{\prime}$09$\farcs$5 & 1645.0 &   278 &   14.8 &  8.1 &    7.6\\
       6 &  12$^{\rm h}$01$^{\rm m}$54$^{\rm s}.$89,$-$18$^{\circ}$52$^{\prime}$51$\farcs$8 & 1449.9 &   273 &   16.1 &  8.1 &    7.7\\
       7 &  12$^{\rm h}$01$^{\rm m}$54$^{\rm s}.$91,$-$18$^{\circ}$52$^{\prime}$52$\farcs$4 & 1469.4 &   311 &   15.6 &  8.1 &    7.7\\
       8 &  12$^{\rm h}$01$^{\rm m}$55$^{\rm s}.$38,$-$18$^{\circ}$52$^{\prime}$48$\farcs$2 & 1391.4 &   282 &   17.6 &  8.2 &    7.8\\
       9 &  12$^{\rm h}$01$^{\rm m}$54$^{\rm s}.$93,$-$18$^{\circ}$53$^{\prime}$02$\farcs$3 & 1557.2 &   317 &   22.5 &  8.4 &    8.1\\
      10 &  12$^{\rm h}$01$^{\rm m}$53$^{\rm s}.$56,$-$18$^{\circ}$53$^{\prime}$08$\farcs$9 & 1596.2 &   184 &   15.7 &  7.9 &    7.5\\
      11 &  12$^{\rm h}$01$^{\rm m}$55$^{\rm s}.$19,$-$18$^{\circ}$52$^{\prime}$40$\farcs$7 & 1576.7 &   275 &   12.3 &  7.9 &    7.5\\
      12 &  12$^{\rm h}$01$^{\rm m}$53$^{\rm s}.$58,$-$18$^{\circ}$53$^{\prime}$08$\farcs$9 & 1610.9 &   142 &    8.6 &  7.4 &    6.9\\
      13 &  12$^{\rm h}$01$^{\rm m}$53$^{\rm s}.$56,$-$18$^{\circ}$53$^{\prime}$09$\farcs$2 & 1625.5 &   209 &   25.4 &  8.0 &    8.0\\
      14 &  12$^{\rm h}$01$^{\rm m}$54$^{\rm s}.$66,$-$18$^{\circ}$53$^{\prime}$05$\farcs$0 & 1474.3 &   293 &   13.1 &  8.0 &    7.5\\
      15 &  12$^{\rm h}$01$^{\rm m}$55$^{\rm s}.$48,$-$18$^{\circ}$52$^{\prime}$46$\farcs$4 & 1459.7 &   240 &   25.6 &  8.0 &    8.0\\
      16 &  12$^{\rm h}$01$^{\rm m}$54$^{\rm s}.$21,$-$18$^{\circ}$53$^{\prime}$08$\farcs$0 & 1425.5 &   226 &   17.7 &  7.7 &    7.7\\
      17 &  12$^{\rm h}$01$^{\rm m}$53$^{\rm s}.$12,$-$18$^{\circ}$53$^{\prime}$09$\farcs$8 & 1703.5 &   241 &   16.5 &  7.8 &    7.7\\
      18 &  12$^{\rm h}$01$^{\rm m}$54$^{\rm s}.$79,$-$18$^{\circ}$52$^{\prime}$54$\farcs$5 & 1610.9 &   149 &   17.9 &  7.6 &    7.5\\
      19 &  12$^{\rm h}$01$^{\rm m}$54$^{\rm s}.$93,$-$18$^{\circ}$52$^{\prime}$58$\farcs$7 & 1493.8 &   262 &   19.2 &  7.6 &    7.8\\
      20 &  12$^{\rm h}$01$^{\rm m}$54$^{\rm s}.$76,$-$18$^{\circ}$52$^{\prime}$55$\farcs$7 & 1645.0 &   183 &   14.9 &  7.6 &    7.5\\
      21 &  12$^{\rm h}$01$^{\rm m}$53$^{\rm s}.$20,$-$18$^{\circ}$53$^{\prime}$16$\farcs$7 & 1727.9 &   224 &   11.9 &  7.5 &    7.3\\
      22 &  12$^{\rm h}$01$^{\rm m}$55$^{\rm s}.$40,$-$18$^{\circ}$53$^{\prime}$02$\farcs$0 & 1518.2 &   168 &   10.4 &  7.4 &    7.1\\
      23 &  12$^{\rm h}$01$^{\rm m}$53$^{\rm s}.$22,$-$18$^{\circ}$53$^{\prime}$16$\farcs$1 & 1708.4 &   249 &    9.1 &  7.4 &    7.2\\
      24 &  12$^{\rm h}$01$^{\rm m}$54$^{\rm s}.$36,$-$18$^{\circ}$53$^{\prime}$08$\farcs$0 & 1454.8 &   170 &   11.6 &  7.5 &    7.2\\
      25 &  12$^{\rm h}$01$^{\rm m}$54$^{\rm s}.$89,$-$18$^{\circ}$53$^{\prime}$06$\farcs$2 & 1610.9 &   244 &   20.5 &  7.7 &    7.9\\
      26 &  12$^{\rm h}$01$^{\rm m}$54$^{\rm s}.$45,$-$18$^{\circ}$52$^{\prime}$49$\farcs$4 & 1479.2 &   118 &   12.2 &  7.1 &    7.1\\
      27 &  12$^{\rm h}$01$^{\rm m}$54$^{\rm s}.$60,$-$18$^{\circ}$52$^{\prime}$45$\farcs$5 & 1601.1 &   225 &    9.9 &  7.4 &    7.2\\
      28 &  12$^{\rm h}$01$^{\rm m}$53$^{\rm s}.$20,$-$18$^{\circ}$53$^{\prime}$16$\farcs$4 & 1679.1 &   266 &   16.4 &  7.5 &    7.7\\
      29 &  12$^{\rm h}$01$^{\rm m}$54$^{\rm s}.$45,$-$18$^{\circ}$52$^{\prime}$44$\farcs$9 & 1620.6 &   161 &   12.4 &  7.2 &    7.2\\
      30 &  12$^{\rm h}$01$^{\rm m}$55$^{\rm s}.$00,$-$18$^{\circ}$52$^{\prime}$53$\farcs$0 & 1562.1 &    89 &   10.1 &  7.0 &    6.8\\
      31 &  12$^{\rm h}$01$^{\rm m}$53$^{\rm s}.$66,$-$18$^{\circ}$53$^{\prime}$12$\farcs$2 & 1571.8 &   141 &    7.7 &  7.0 &    6.8\\
      32 &  12$^{\rm h}$01$^{\rm m}$53$^{\rm s}.$64,$-$18$^{\circ}$53$^{\prime}$11$\farcs$9 & 1581.6 &   119 &   11.9 &  7.0 &    7.1\\
      33 &  12$^{\rm h}$01$^{\rm m}$54$^{\rm s}.$89,$-$18$^{\circ}$52$^{\prime}$45$\farcs$5 & 1381.6 &   118 &    9.1 &  7.0 &    6.8\\
      34 &  12$^{\rm h}$01$^{\rm m}$55$^{\rm s}.$65,$-$18$^{\circ}$52$^{\prime}$51$\farcs$5 & 1537.7 &    79 &    5.6 &  6.7 &    6.2\\
      35 &  12$^{\rm h}$01$^{\rm m}$53$^{\rm s}.$16,$-$18$^{\circ}$53$^{\prime}$20$\farcs$9 & 1762.0 &   113 &    9.5 &  7.0 &    6.9\\
      36 &  12$^{\rm h}$01$^{\rm m}$54$^{\rm s}.$55,$-$18$^{\circ}$52$^{\prime}$45$\farcs$2 & 1615.7 &    94 &   12.9 &  7.0 &    7.0\\
      37 &  12$^{\rm h}$01$^{\rm m}$54$^{\rm s}.$91,$-$18$^{\circ}$53$^{\prime}$07$\farcs$1 & 1645.0 &   216 &   10.4 &  7.3 &    7.2\\
      38 &  12$^{\rm h}$01$^{\rm m}$53$^{\rm s}.$92,$-$18$^{\circ}$53$^{\prime}$18$\farcs$2 & 1703.5 &   211 &    7.7 &  7.0 &    6.9\\
      39 &  12$^{\rm h}$01$^{\rm m}$54$^{\rm s}.$40,$-$18$^{\circ}$52$^{\prime}$49$\farcs$7 & 1488.9 &    55 &   10.7 &  6.9 &    6.6\\
      40 &  12$^{\rm h}$01$^{\rm m}$54$^{\rm s}.$87,$-$18$^{\circ}$52$^{\prime}$59$\farcs$6 & 1610.9 &   204 &   14.4 &  7.2 &    7.5\\
      41 &  12$^{\rm h}$01$^{\rm m}$54$^{\rm s}.$81,$-$18$^{\circ}$52$^{\prime}$47$\farcs$6 & 1386.5 &   261 &   18.9 &  7.4 &    7.8\\
      42 &  12$^{\rm h}$01$^{\rm m}$54$^{\rm s}.$11,$-$18$^{\circ}$53$^{\prime}$26$\farcs$3 & 1606.0 &    53 &    4.2 &  6.4 &    5.8\\
      43 &  12$^{\rm h}$01$^{\rm m}$55$^{\rm s}.$23,$-$18$^{\circ}$53$^{\prime}$09$\farcs$5 & 1630.4 &   168 &    8.7 &  6.8 &    6.9\\
      44 &  12$^{\rm h}$01$^{\rm m}$54$^{\rm s}.$00,$-$18$^{\circ}$53$^{\prime}$09$\farcs$5 & 1498.7 &    70 &    7.3 &  6.7 &    6.4\\
      45 &  12$^{\rm h}$01$^{\rm m}$54$^{\rm s}.$57,$-$18$^{\circ}$52$^{\prime}$46$\farcs$4 & 1430.4 &   229 &   11.7 &  7.2 &    7.3\\
      46 &  12$^{\rm h}$01$^{\rm m}$55$^{\rm s}.$25,$-$18$^{\circ}$53$^{\prime}$09$\farcs$2 & 1640.1 &    75 &    7.2 &  6.6 &    6.4\\
      47 &  12$^{\rm h}$01$^{\rm m}$54$^{\rm s}.$74,$-$18$^{\circ}$52$^{\prime}$40$\farcs$4 & 1513.3 & $<$80 &   12.1 &  6.6 & $<$6.9\\
      48 &  12$^{\rm h}$01$^{\rm m}$54$^{\rm s}.$28,$-$18$^{\circ}$52$^{\prime}$50$\farcs$3 & 1532.8 &   134 &   11.9 &  6.9 &    7.1\\
      49 &  12$^{\rm h}$01$^{\rm m}$54$^{\rm s}.$55,$-$18$^{\circ}$52$^{\prime}$45$\farcs$8 & 1396.3 &    71 &    5.9 &  6.5 &    6.2\\
      50 &  12$^{\rm h}$01$^{\rm m}$55$^{\rm s}.$33,$-$18$^{\circ}$52$^{\prime}$43$\farcs$7 & 1645.0 &   176 &   12.0 &  6.8 &    7.2\\
      51 &  12$^{\rm h}$01$^{\rm m}$54$^{\rm s}.$76,$-$18$^{\circ}$52$^{\prime}$53$\farcs$3 & 1527.9 &   102 &   16.0 &  6.8 &    7.3\\
      52 &  12$^{\rm h}$01$^{\rm m}$55$^{\rm s}.$48,$-$18$^{\circ}$53$^{\prime}$07$\farcs$4 & 1601.1 &   149 &   10.0 &  6.8 &    7.0\\
      53 &  12$^{\rm h}$01$^{\rm m}$53$^{\rm s}.$66,$-$18$^{\circ}$53$^{\prime}$12$\farcs$8 & 1552.3 &    97 &    7.0 &  6.5 &    6.5\\
      54 &  12$^{\rm h}$01$^{\rm m}$55$^{\rm s}.$02,$-$18$^{\circ}$52$^{\prime}$52$\farcs$7 & 1552.3 &   227 &   18.1 &  7.1 &    7.7\\
      55 &  12$^{\rm h}$01$^{\rm m}$54$^{\rm s}.$17,$-$18$^{\circ}$52$^{\prime}$47$\farcs$3 & 1513.3 &   109 &    7.1 &  6.6 &    6.6\\
      56 &  12$^{\rm h}$01$^{\rm m}$55$^{\rm s}.$25,$-$18$^{\circ}$52$^{\prime}$43$\farcs$1 & 1537.7 &   114 &   10.0 &  6.6 &    6.9\\
      57 &  12$^{\rm h}$01$^{\rm m}$54$^{\rm s}.$26,$-$18$^{\circ}$52$^{\prime}$49$\farcs$4 & 1523.1 &   127 &   12.8 &  6.8 &    7.2\\
      58 &  12$^{\rm h}$01$^{\rm m}$54$^{\rm s}.$17,$-$18$^{\circ}$52$^{\prime}$45$\farcs$8 & 1615.7 & $<$80 &    2.1 &  6.3 & $<$5.4\\
      59 &  12$^{\rm h}$01$^{\rm m}$54$^{\rm s}.$49,$-$18$^{\circ}$52$^{\prime}$45$\farcs$2 & 1406.0 &   128 &   10.6 &  6.6 &    7.0\\
      60 &  12$^{\rm h}$01$^{\rm m}$53$^{\rm s}.$64,$-$18$^{\circ}$53$^{\prime}$08$\farcs$9 & 1527.9 &   238 &   11.8 &  7.1 &    7.4\\
      61 &  12$^{\rm h}$01$^{\rm m}$54$^{\rm s}.$19,$-$18$^{\circ}$52$^{\prime}$44$\farcs$6 & 1610.9 & $<$80 &    9.4 &  6.3 & $<$6.7\\
      62 &  12$^{\rm h}$01$^{\rm m}$53$^{\rm s}.$94,$-$18$^{\circ}$53$^{\prime}$17$\farcs$9 & 1684.0 &   220 &    6.2 &  6.8 &    6.8\\
      63 &  12$^{\rm h}$01$^{\rm m}$54$^{\rm s}.$66,$-$18$^{\circ}$52$^{\prime}$41$\farcs$3 & 1518.2 &    35 &    7.4 &  6.6 &    6.1\\
      64 &  12$^{\rm h}$01$^{\rm m}$53$^{\rm s}.$28,$-$18$^{\circ}$53$^{\prime}$21$\farcs$8 & 1737.7 & $<$80 &   11.2 &  6.3 & $<$6.8\\
      65 &  12$^{\rm h}$01$^{\rm m}$55$^{\rm s}.$65,$-$18$^{\circ}$52$^{\prime}$41$\farcs$9 & 1606.0 &   144 &   11.0 &  6.5 &    7.1\\
      66 &  12$^{\rm h}$01$^{\rm m}$54$^{\rm s}.$38,$-$18$^{\circ}$52$^{\prime}$47$\farcs$3 & 1527.9 &   163 &   13.9 &  6.9 &    7.3\\
      67 &  12$^{\rm h}$01$^{\rm m}$53$^{\rm s}.$24,$-$18$^{\circ}$53$^{\prime}$17$\farcs$0 & 1762.0 &   189 &   10.4 &  6.8 &    7.2\\
      68 &  12$^{\rm h}$01$^{\rm m}$53$^{\rm s}.$69,$-$18$^{\circ}$53$^{\prime}$13$\farcs$4 & 1645.0 &   166 &   24.0 &  6.8 &    7.8\\
      69 &  12$^{\rm h}$01$^{\rm m}$54$^{\rm s}.$17,$-$18$^{\circ}$52$^{\prime}$45$\farcs$8 & 1601.1 & $<$80 &    6.3 &  6.1 & $<$6.3\\
      70 &  12$^{\rm h}$01$^{\rm m}$54$^{\rm s}.$72,$-$18$^{\circ}$52$^{\prime}$53$\farcs$3 & 1547.5 &    99 &   10.7 &  6.7 &    6.9\\
      71 &  12$^{\rm h}$01$^{\rm m}$54$^{\rm s}.$13,$-$18$^{\circ}$52$^{\prime}$47$\farcs$3 & 1459.7 & $<$80 & $<$4.9 &  5.4 & $<$6.1\\
      72 &  12$^{\rm h}$01$^{\rm m}$54$^{\rm s}.$17,$-$18$^{\circ}$53$^{\prime}$17$\farcs$0 & 1532.8 &   130 &    9.0 &  6.5 &    6.9\\
      73 &  12$^{\rm h}$01$^{\rm m}$55$^{\rm s}.$50,$-$18$^{\circ}$52$^{\prime}$52$\farcs$4 & 1474.3 & $<$80 &    1.6 &  5.9 & $<$5.1\\
      74 &  12$^{\rm h}$01$^{\rm m}$53$^{\rm s}.$07,$-$18$^{\circ}$53$^{\prime}$18$\farcs$2 & 1445.0 & $<$80 & $<$4.9 &  5.6 & $<$6.1\\
      75 &  12$^{\rm h}$01$^{\rm m}$55$^{\rm s}.$59,$-$18$^{\circ}$52$^{\prime}$53$\farcs$3 & 1493.8 & $<$80 &    2.0 &  5.9 & $<$5.3\\
      76 &  12$^{\rm h}$01$^{\rm m}$55$^{\rm s}.$50,$-$18$^{\circ}$52$^{\prime}$52$\farcs$7 & 1440.2 & $<$80 &    3.5 &  5.9 & $<$5.8\\
      77 &  12$^{\rm h}$01$^{\rm m}$53$^{\rm s}.$73,$-$18$^{\circ}$53$^{\prime}$13$\farcs$1 & 1664.5 & $<$80 &    2.0 &  5.8 & $<$5.4\\
      78 &  12$^{\rm h}$01$^{\rm m}$54$^{\rm s}.$21,$-$18$^{\circ}$52$^{\prime}$43$\farcs$7 & 1415.8 & $<$80 & $<$4.9 &  5.2 & $<$6.1\\
      79 &  12$^{\rm h}$01$^{\rm m}$53$^{\rm s}.$60,$-$18$^{\circ}$53$^{\prime}$17$\farcs$6 & 1430.4 & $<$80 & $<$4.9 &  5.5 & $<$6.1\\
      80 &  12$^{\rm h}$01$^{\rm m}$54$^{\rm s}.$05,$-$18$^{\circ}$53$^{\prime}$18$\farcs$2 & 1498.7 & $<$80 &    4.2 &  5.9 & $<$6.0\\
      81 &  12$^{\rm h}$01$^{\rm m}$55$^{\rm s}.$38,$-$18$^{\circ}$53$^{\prime}$05$\farcs$6 & 1567.0 & $<$80 &    2.5 &  5.9 & $<$5.5\\
      82 &  12$^{\rm h}$01$^{\rm m}$54$^{\rm s}.$30,$-$18$^{\circ}$52$^{\prime}$41$\farcs$9 & 1518.2 & $<$80 &    1.2 &  5.7 & $<$4.9\\
      83 &  12$^{\rm h}$01$^{\rm m}$53$^{\rm s}.$16,$-$18$^{\circ}$53$^{\prime}$09$\farcs$5 & 1479.2 &   105 & $<$4.9 &  6.1 & $<$6.2\\
      84 &  12$^{\rm h}$01$^{\rm m}$54$^{\rm s}.$26,$-$18$^{\circ}$52$^{\prime}$54$\farcs$2 & 1659.6 & $<$80 & $<$4.9 &  5.4 & $<$6.1\\
      85 &  12$^{\rm h}$01$^{\rm m}$54$^{\rm s}.$40,$-$18$^{\circ}$52$^{\prime}$44$\farcs$3 & 1376.8 & $<$80 & $<$4.9 &  5.8 & $<$6.1\\
      86 &  12$^{\rm h}$01$^{\rm m}$55$^{\rm s}.$61,$-$18$^{\circ}$52$^{\prime}$51$\farcs$5 & 1508.4 & $<$80 & $<$4.9 &  5.4 & $<$6.1\\
      87 &  12$^{\rm h}$01$^{\rm m}$53$^{\rm s}.$18,$-$18$^{\circ}$53$^{\prime}$15$\farcs$5 & 1396.3 & $<$80 & $<$4.9 &  5.3 & $<$6.1\\
      88 &  12$^{\rm h}$01$^{\rm m}$54$^{\rm s}.$53,$-$18$^{\circ}$53$^{\prime}$24$\farcs$8 & 1513.3 &    41 &    2.4 &  6.0 &    5.2\\
      89 &  12$^{\rm h}$01$^{\rm m}$55$^{\rm s}.$23,$-$18$^{\circ}$52$^{\prime}$41$\farcs$0 & 1615.7 & $<$80 & $<$4.9 &  5.2 & $<$6.1\\
      90 &  12$^{\rm h}$01$^{\rm m}$53$^{\rm s}.$69,$-$18$^{\circ}$53$^{\prime}$10$\farcs$7 & 1488.9 & $<$80 & $<$4.9 &  5.5 & $<$6.1\\
      91 &  12$^{\rm h}$01$^{\rm m}$54$^{\rm s}.$24,$-$18$^{\circ}$52$^{\prime}$45$\farcs$2 & 1367.0 & $<$80 & $<$4.9 &  5.3 & $<$6.1\\
      92 &  12$^{\rm h}$01$^{\rm m}$54$^{\rm s}.$53,$-$18$^{\circ}$53$^{\prime}$17$\farcs$3 & 1674.3 & $<$80 &    1.8 &  5.7 & $<$5.2\\
      93 &  12$^{\rm h}$01$^{\rm m}$54$^{\rm s}.$11,$-$18$^{\circ}$53$^{\prime}$26$\farcs$6 & 1615.7 & $<$80 & $<$4.9 &  5.4 & $<$6.1\\
      94 &  12$^{\rm h}$01$^{\rm m}$53$^{\rm s}.$88,$-$18$^{\circ}$53$^{\prime}$08$\farcs$3 & 1479.2 & $<$80 &    0.8 &  5.6 & $<$4.5\\
      95 &  12$^{\rm h}$01$^{\rm m}$54$^{\rm s}.$85,$-$18$^{\circ}$53$^{\prime}$16$\farcs$1 & 1664.5 & $<$80 & $<$4.9 &  5.5 & $<$6.1\\
      96 &  12$^{\rm h}$01$^{\rm m}$53$^{\rm s}.$71,$-$18$^{\circ}$53$^{\prime}$13$\farcs$7 & 1527.9 & $<$80 & $<$4.9 &  5.2 & $<$6.1\\
      97 &  12$^{\rm h}$01$^{\rm m}$53$^{\rm s}.$20,$-$18$^{\circ}$53$^{\prime}$08$\farcs$3 & 1464.5 & $<$80 & $<$4.9 &  5.6 & $<$6.1\\
      98 &  12$^{\rm h}$01$^{\rm m}$53$^{\rm s}.$58,$-$18$^{\circ}$53$^{\prime}$10$\farcs$1 & 1732.8 & $<$80 & $<$4.9 &  5.1 & $<$6.1\\
      99 &  12$^{\rm h}$01$^{\rm m}$54$^{\rm s}.$11,$-$18$^{\circ}$52$^{\prime}$49$\farcs$1 & 1606.0 & $<$80 & $<$4.9 &  5.3 & $<$6.1\\
     100 &  12$^{\rm h}$01$^{\rm m}$54$^{\rm s}.$11,$-$18$^{\circ}$53$^{\prime}$26$\farcs$3 & 1649.9 & $<$80 &    1.0 &  5.4 & $<$4.7\\
     101 &  12$^{\rm h}$01$^{\rm m}$54$^{\rm s}.$34,$-$18$^{\circ}$52$^{\prime}$47$\farcs$9 & 1567.0 & $<$80 & $<$4.9 &  5.6 & $<$6.1\\
     102 &  12$^{\rm h}$01$^{\rm m}$54$^{\rm s}.$19,$-$18$^{\circ}$53$^{\prime}$18$\farcs$5 & 1503.6 & $<$80 & $<$4.9 &  5.3 & $<$6.1\\
     103 &  12$^{\rm h}$01$^{\rm m}$54$^{\rm s}.$91,$-$18$^{\circ}$52$^{\prime}$38$\farcs$9 & 1328.0 & $<$80 & $<$4.9 &  5.1 & $<$6.1\\
     104 &  12$^{\rm h}$01$^{\rm m}$53$^{\rm s}.$79,$-$18$^{\circ}$53$^{\prime}$20$\farcs$6 & 1484.1 & $<$80 &    4.5 &  5.7 & $<$6.0\\
     105 &  12$^{\rm h}$01$^{\rm m}$53$^{\rm s}.$73,$-$18$^{\circ}$53$^{\prime}$19$\farcs$7 & 1727.9 & $<$80 & $<$4.9 &  5.1 & $<$6.1\\
     106 &  12$^{\rm h}$01$^{\rm m}$54$^{\rm s}.$13,$-$18$^{\circ}$53$^{\prime}$26$\farcs$9 & 1635.2 & $<$80 & $<$4.9 &  4.9 & $<$6.1\\
     107 &  12$^{\rm h}$01$^{\rm m}$54$^{\rm s}.$28,$-$18$^{\circ}$52$^{\prime}$51$\farcs$5 & 1669.4 & $<$80 & $<$4.9 &  5.3 & $<$6.1\\
     108 &  12$^{\rm h}$01$^{\rm m}$54$^{\rm s}.$21,$-$18$^{\circ}$53$^{\prime}$11$\farcs$6 & 1537.7 & $<$80 & $<$4.9 &  5.2 & $<$6.1\\
     109 &  12$^{\rm h}$01$^{\rm m}$54$^{\rm s}.$17,$-$18$^{\circ}$52$^{\prime}$49$\farcs$7 & 1479.2 & $<$80 & $<$4.9 &  5.3 & $<$6.1\\
     110 &  12$^{\rm h}$01$^{\rm m}$55$^{\rm s}.$44,$-$18$^{\circ}$52$^{\prime}$45$\farcs$8 & 1645.0 & $<$80 & $<$4.9 &  5.4 & $<$6.1\\
     111 &  12$^{\rm h}$01$^{\rm m}$53$^{\rm s}.$22,$-$18$^{\circ}$53$^{\prime}$12$\farcs$8 & 1488.9 & $<$80 & $<$4.9 &  5.0 & $<$6.1\\
     112 &  12$^{\rm h}$01$^{\rm m}$54$^{\rm s}.$02,$-$18$^{\circ}$53$^{\prime}$16$\farcs$4 & 1742.5 & $<$80 & $<$4.9 &  5.5 & $<$6.1\\
     113 &  12$^{\rm h}$01$^{\rm m}$54$^{\rm s}.$21,$-$18$^{\circ}$53$^{\prime}$21$\farcs$8 & 1732.8 & $<$80 & $<$4.9 &  5.2 & $<$6.1\\
     114 &  12$^{\rm h}$01$^{\rm m}$54$^{\rm s}.$28,$-$18$^{\circ}$52$^{\prime}$43$\farcs$4 & 1537.7 & $<$80 & $<$4.9 &  5.1 & $<$6.1\\
     115 &  12$^{\rm h}$01$^{\rm m}$54$^{\rm s}.$72,$-$18$^{\circ}$53$^{\prime}$05$\farcs$9 & 1654.8 & $<$80 & $<$4.9 &  5.5 & $<$6.1\\
     116 &  12$^{\rm h}$01$^{\rm m}$53$^{\rm s}.$18,$-$18$^{\circ}$53$^{\prime}$09$\farcs$2 & 1493.8 & $<$80 &    1.3 &  5.3 & $<$4.9\\
     117 &  12$^{\rm h}$01$^{\rm m}$54$^{\rm s}.$62,$-$18$^{\circ}$53$^{\prime}$08$\farcs$0 & 1381.6 & $<$80 & $<$4.9 &  5.0 & $<$6.1\\
     118 &  12$^{\rm h}$01$^{\rm m}$53$^{\rm s}.$88,$-$18$^{\circ}$53$^{\prime}$11$\farcs$6 & 1527.9 & $<$80 & $<$4.9 &  5.0 & $<$6.1\\
     119 &  12$^{\rm h}$01$^{\rm m}$54$^{\rm s}.$26,$-$18$^{\circ}$52$^{\prime}$44$\farcs$0 & 1547.5 & $<$80 &    1.0 &  4.9 & $<$4.8\\
     120 &  12$^{\rm h}$01$^{\rm m}$54$^{\rm s}.$98,$-$18$^{\circ}$52$^{\prime}$44$\farcs$3 & 1645.0 & $<$80 & $<$4.9 &  5.3 & $<$6.1\\
     121 &  12$^{\rm h}$01$^{\rm m}$53$^{\rm s}.$43,$-$18$^{\circ}$53$^{\prime}$18$\farcs$8 & 1391.4 & $<$80 & $<$4.9 &  4.9 & $<$6.1\\
     122 &  12$^{\rm h}$01$^{\rm m}$53$^{\rm s}.$69,$-$18$^{\circ}$53$^{\prime}$17$\farcs$6 & 1718.2 & $<$80 & $<$4.9 &  4.9 & $<$6.1\\
     123 &  12$^{\rm h}$01$^{\rm m}$55$^{\rm s}.$48,$-$18$^{\circ}$52$^{\prime}$45$\farcs$5 & 1620.6 & $<$80 & $<$4.9 &  5.1 & $<$6.1\\
     124 &  12$^{\rm h}$01$^{\rm m}$54$^{\rm s}.$19,$-$18$^{\circ}$52$^{\prime}$47$\farcs$6 & 1674.3 & $<$80 & $<$4.9 &  4.9 & $<$6.1\\
     125 &  12$^{\rm h}$01$^{\rm m}$53$^{\rm s}.$54,$-$18$^{\circ}$53$^{\prime}$18$\farcs$2 & 1786.4 & $<$80 & $<$4.9 &  4.9 & $<$6.1\\
     126 &  12$^{\rm h}$01$^{\rm m}$55$^{\rm s}.$61,$-$18$^{\circ}$52$^{\prime}$43$\farcs$7 & 1586.5 & $<$80 & $<$4.9 &  4.9 & $<$6.1\\
     127 &  12$^{\rm h}$01$^{\rm m}$54$^{\rm s}.$11,$-$18$^{\circ}$53$^{\prime}$26$\farcs$9 & 1537.7 & $<$80 & $<$4.9 &  4.9 & $<$6.1\\
     128 &  12$^{\rm h}$01$^{\rm m}$54$^{\rm s}.$28,$-$18$^{\circ}$53$^{\prime}$16$\farcs$1 & 1562.1 & $<$80 & $<$4.9 &  4.9 & $<$6.1\\
     129 &  12$^{\rm h}$01$^{\rm m}$53$^{\rm s}.$92,$-$18$^{\circ}$53$^{\prime}$06$\farcs$5 & 1513.3 & $<$80 & $<$4.9 &  4.9 & $<$6.1\\
     130 &  12$^{\rm h}$01$^{\rm m}$54$^{\rm s}.$09,$-$18$^{\circ}$53$^{\prime}$25$\farcs$1 & 1635.2 & $<$80 & $<$4.9 &  5.1 & $<$6.1\\
     131 &  12$^{\rm h}$01$^{\rm m}$54$^{\rm s}.$15,$-$18$^{\circ}$53$^{\prime}$10$\farcs$7 & 1640.1 & $<$80 & $<$4.9 &  5.0 & $<$6.1\\
     132 &  12$^{\rm h}$01$^{\rm m}$54$^{\rm s}.$74,$-$18$^{\circ}$53$^{\prime}$05$\farcs$9 & 1674.3 & $<$80 &    1.3 &  4.9 & $<$4.9
  \enddata
  
\tablecomments{Basic cloud properties from {\tt clumpfind}. Clouds
  that were unresolved spatially or in velocity are assigned upper
  limits of 80\,pc and 4.9\,km\,s$^{-1}$, respectively. Note that the
  velocity is defined using the radio definition
  ($v_{\rm}\,=\,cz/(1+z)$).}

\end{deluxetable*}


\begin{thebibliography}{80}
\expandafter\ifx\csname natexlab\endcsname\relax\def\natexlab#1{#1}\fi

\bibitem[{{Ashman} \& {Zepf}(2001)}]{ashman01}
{Ashman}, K.~M., \& {Zepf}, S.~E. 2001, \aj, 122, 1888

\bibitem[{{Barth} {et~al.}(1995){Barth}, {Ho}, {Filippenko}, \&
  {Sargent}}]{barth95}
{Barth}, A.~J., {Ho}, L.~C., {Filippenko}, A.~V., \& {Sargent}, W.~L. 1995,
  \aj, 110, 1009

\bibitem[{{Bastian} {et~al.}(2009){Bastian}, {Trancho}, {Konstantopoulos}, \&
  {Miller}}]{bastian09}
{Bastian}, N., {Trancho}, G., {Konstantopoulos}, I.~S., \& {Miller}, B.~W.
  2009, \apj, 701, 607

\bibitem[{{Bodenheimer} \& {Sweigart}(1968)}]{bodenheimer68}
{Bodenheimer}, P., \& {Sweigart}, A. 1968, \apj, 152, 515

\bibitem[{{Bolatto} {et~al.}(2008){Bolatto}, {Leroy}, {Rosolowsky}, {Walter},
  \& {Blitz}}]{bolatto08}
{Bolatto}, A.~D., {Leroy}, A.~K., {Rosolowsky}, E., {Walter}, F., \& {Blitz},
  L. 2008, \apj, 686, 948

\bibitem[{{Braine} {et~al.}(1993){Braine}, {Combes}, {Casoli}, {Dupraz},
  {Gerin}, {Klein}, {Wielebinski}, \& {Brouillet}}]{braine93}
{Braine}, J., {Combes}, F., {Casoli}, F., {Dupraz}, C., {Gerin}, M., {Klein},
  U., {Wielebinski}, R., \& {Brouillet}, N. 1993, \aaps, 97, 887

\bibitem[{{Brandl} {et~al.}(2009){Brandl}, {Snijders}, {den Brok}, {Whelan},
  {Groves}, {van der Werf}, {Charmandaris}, {Smith}, {Armus}, {Kennicutt}, \&
  {Houck}}]{brandl09}
{Brandl}, B.~R., {et~al.} 2009, \apj, 699, 1982

\bibitem[{{Brogan} {et~al.}(2010){Brogan}, {Johnson}, \& {Darling}}]{brogan10}
{Brogan}, C., {Johnson}, K., \& {Darling}, J. 2010, \apjl, 716, L51

\bibitem[{{Cen}(2001)}]{cen01}
{Cen}, R. 2001, \apj, 560, 592

\bibitem[{{Elmegreen} \& {Efremov}(1997)}]{elmegreen97}
{Elmegreen}, B.~G., \& {Efremov}, Y.~N. 1997, \apj, 480, 235

\bibitem[{{Evans} {et~al.}(1997){Evans}, {Harper}, \& {Helou}}]{evans97}
{Evans}, R., {Harper}, A., \& {Helou}, G. 1997, in Extragalactic Astronomy in
  the Infrared, ed. {G.~A.~Mamon, T.~X.~Thuan, \& J.~Tran Thanh Van}, 143

\bibitem[{{Fabbiano} {et~al.}(2003){Fabbiano}, {Zezas}, {King}, {Ponman},
  {Rots}, \& {Schweizer}}]{fabbiano03}
{Fabbiano}, G., {Zezas}, A., {King}, A.~R., {Ponman}, T.~J., {Rots}, A., \&
  {Schweizer}, F. 2003, \apjl, 584, L5

\bibitem[{{Fabbiano} {et~al.}(2001){Fabbiano}, {Zezas}, \&
  {Murray}}]{fabbiano01}
{Fabbiano}, G., {Zezas}, A., \& {Murray}, S.~S. 2001, \apj, 554, 1035

\bibitem[{{Fall} {et~al.}(2005){Fall}, {Chandar}, \& {Whitmore}}]{fall05}
{Fall}, S.~M., {Chandar}, R., \& {Whitmore}, B.~C. 2005, \apjl, 631, L133

\bibitem[{{Feigelson} \& {Nelson}(1985)}]{feigelson85}
{Feigelson}, E.~D., \& {Nelson}, P.~I. 1985, \apj, 293, 192

\bibitem[{{Gao}(2008)}]{gao08}
{Gao}, Y. 2008, \nat, 452, 417

\bibitem[{{Gao} {et~al.}(2007){Gao}, {Carilli}, {Solomon}, \& {Vanden
  Bout}}]{gao07}
{Gao}, Y., {Carilli}, C.~L., {Solomon}, P.~M., \& {Vanden Bout}, P.~A. 2007,
  \apjl, 660, L93

\bibitem[{{Gao} {et~al.}(2001){Gao}, {Lo}, {Lee}, \& {Lee}}]{gao01}
{Gao}, Y., {Lo}, K.~Y., {Lee}, S.-W., \& {Lee}, T.-H. 2001, \apj, 548, 172

\bibitem[{{Gao} \& {Solomon}(2004)}]{gao04}
{Gao}, Y., \& {Solomon}, P.~M. 2004, \apj, 606, 271

\bibitem[{{Harris} \& {Pudritz}(1994)}]{harris94}
{Harris}, W.~E., \& {Pudritz}, R.~E. 1994, \apj, 429, 177

\bibitem[{{Herrera} {et~al.}(2011){Herrera}, {Boulanger}, \&
  {Nesvadba}}]{herrera11}
{Herrera}, C.~N., {Boulanger}, F., \& {Nesvadba}, N.~P.~H. 2011, \aap, 534,
  A138

\bibitem[{{Heyer} {et~al.}(2009){Heyer}, {Krawczyk}, {Duval}, \&
  {Jackson}}]{heyer09}
{Heyer}, M., {Krawczyk}, C., {Duval}, J., \& {Jackson}, J.~M. 2009, \apj, 699,
  1092

\bibitem[{{Hibbard} {et~al.}(2001){Hibbard}, {van der Hulst}, {Barnes}, \&
  {Rich}}]{hibbard01}
{Hibbard}, J.~E., {van der Hulst}, J.~M., {Barnes}, J.~E., \& {Rich}, R.~M.
  2001, \aj, 122, 2969

\bibitem[{{Hibbard} {et~al.}(2005){Hibbard}, {Bianchi}, {Thilker}, {Rich},
  {Schiminovich}, {Xu}, {Neff}, {Seibert}, {Lauger}, {Burgarella}, {Barlow},
  {Byun}, {Donas}, {Forster}, {Friedman}, {Heckman}, {Jelinsky}, {Lee},
  {Madore}, {Malina}, {Martin}, {Milliard}, {Morrissey}, {Siegmund}, {Small},
  {Szalay}, {Welsh}, \& {Wyder}}]{hibbard05}
{Hibbard}, J.~E., {et~al.} 2005, \apjl, 619, L87

\bibitem[{{Ho}(1997)}]{ho97}
{Ho}, L.~C. 1997, in Revista Mexicana de Astronomia y Astrofisica, vol. 27,
  Vol.~6, Revista Mexicana de Astronomia y Astrofisica Conference Series, ed.
  {J.~Franco, R.~Terlevich, \& A.~Serrano}, 5

\bibitem[{{Ho} \& {Filippenko}(1996{\natexlab{a}})}]{ho96a}
{Ho}, L.~C., \& {Filippenko}, A.~V. 1996{\natexlab{a}}, \apjl, 466, L83

\bibitem[{{Ho} \& {Filippenko}(1996{\natexlab{b}})}]{ho96b}
---. 1996{\natexlab{b}}, \apj, 472, 600

\bibitem[{{Ho} {et~al.}(2004){Ho}, {Moran}, \& {Lo}}]{ho04}
{Ho}, P.~T.~P., {Moran}, J.~M., \& {Lo}, K.~Y. 2004, \apjl, 616, L1

\bibitem[{{Holtzman} {et~al.}(1992){Holtzman}, {Faber}, {Shaya}, {Lauer},
  {Groth}, {Hunter}, {Baum}, {Ewald}, {Hester}, {Light}, {Lynds}, {O'Neil}, \&
  {Westphal}}]{holtzman92}
{Holtzman}, J.~A., {et~al.} 1992, \aj, 103, 691

\bibitem[{{Jog} \& {Solomon}(1992)}]{jog92}
{Jog}, C.~J., \& {Solomon}, P.~M. 1992, \apj, 387, 152

\bibitem[{{Keto} \& {Caselli}(2010)}]{keto10}
{Keto}, E., \& {Caselli}, P. 2010, \mnras, 402, 1625

\bibitem[{{Keto} {et~al.}(2005){Keto}, {Ho}, \& {Lo}}]{keto05}
{Keto}, E., {Ho}, L.~C., \& {Lo}, K.-Y. 2005, \apj, 635, 1062

\bibitem[{{Kilgard} {et~al.}(2002){Kilgard}, {Kaaret}, {Krauss}, {Prestwich},
  {Raley}, \& {Zezas}}]{kilgard02}
{Kilgard}, R.~E., {Kaaret}, P., {Krauss}, M.~I., {Prestwich}, A.~H., {Raley},
  M.~T., \& {Zezas}, A. 2002, \apj, 573, 138

\bibitem[{{Klaas} {et~al.}(2010){Klaas}, {Nielbock}, {Haas}, {Krause}, \&
  {Schreiber}}]{klaas10}
{Klaas}, U., {Nielbock}, M., {Haas}, M., {Krause}, O., \& {Schreiber}, J. 2010,
  \aap, 518, L44

\bibitem[{{Larsen}(2010)}]{larsen10}
{Larsen}, S.~S. 2010, Royal Society of London Philosophical Transactions Series
  A, 368, 867

\bibitem[{{Larsen} \& {Richtler}(2004)}]{larsen04}
{Larsen}, S.~S., \& {Richtler}, T. 2004, \aap, 427, 495

\bibitem[{{Larson}(1981)}]{larson81}
{Larson}, R.~B. 1981, \mnras, 194, 809

\bibitem[{{Lupton} {et~al.}(2004){Lupton}, {Blanton}, {Fekete}, {Hogg},
  {O'Mullane}, {Szalay}, \& {Wherry}}]{lupton04}
{Lupton}, R., {Blanton}, M.~R., {Fekete}, G., {Hogg}, D.~W., {O'Mullane}, W.,
  {Szalay}, A., \& {Wherry}, N. 2004, \pasp, 116, 133

\bibitem[{{Maoz} {et~al.}(1996){Maoz}, {Barth}, {Sternberg}, {Filippenko},
  {Ho}, {Macchetto}, {Rix}, \& {Schneider}}]{maoz96}
{Maoz}, D., {Barth}, A.~J., {Sternberg}, A., {Filippenko}, A.~V., {Ho}, L.~C.,
  {Macchetto}, F.~D., {Rix}, H.-W., \& {Schneider}, D.~P. 1996, \aj, 111, 2248

\bibitem[{{McCrady} \& {Graham}(2007)}]{mccrady07}
{McCrady}, N., \& {Graham}, J.~R. 2007, \apj, 663, 844

\bibitem[{{Melo} {et~al.}(2005){Melo}, {Mu{\~n}oz-Tu{\~n}{\'o}n},
  {Ma{\'{\i}}z-Apell{\'a}niz}, \& {Tenorio-Tagle}}]{melo05}
{Melo}, V.~P., {Mu{\~n}oz-Tu{\~n}{\'o}n}, C., {Ma{\'{\i}}z-Apell{\'a}niz}, J.,
  \& {Tenorio-Tagle}, G. 2005, \apj, 619, 270

\bibitem[{{Mengel} {et~al.}(2002){Mengel}, {Lehnert}, {Thatte}, \&
  {Genzel}}]{mengel02}
{Mengel}, S., {Lehnert}, M.~D., {Thatte}, N., \& {Genzel}, R. 2002, \aap, 383,
  137

\bibitem[{{Mengel} {et~al.}(2005){Mengel}, {Lehnert}, {Thatte}, \&
  {Genzel}}]{mengel05}
---. 2005, \aap, 443, 41

\bibitem[{{Meurer} {et~al.}(1995){Meurer}, {Heckman}, {Leitherer}, {Kinney},
  {Robert}, \& {Garnett}}]{meurer95}
{Meurer}, G.~R., {Heckman}, T.~M., {Leitherer}, C., {Kinney}, A., {Robert}, C.,
  \& {Garnett}, D.~R. 1995, \aj, 110, 2665

\bibitem[{{Mirabel} {et~al.}(1998){Mirabel}, {Vigroux}, {Charmandaris},
  {Sauvage}, {Gallais}, {Tran}, {Cesarsky}, {Madden}, \& {Duc}}]{mirabel98}
{Mirabel}, I.~F., {et~al.} 1998, \aap, 333, L1

\bibitem[{{Neff} \& {Ulvestad}(2000)}]{neff00}
{Neff}, S.~G., \& {Ulvestad}, J.~S. 2000, \aj, 120, 670

\bibitem[{{Oka} {et~al.}(1998){Oka}, {Hasegawa}, {Hayashi}, {Handa}, \&
  {Sakamoto}}]{oka98}
{Oka}, T., {Hasegawa}, T., {Hayashi}, M., {Handa}, T., \& {Sakamoto}, S. 1998,
  \apj, 493, 730

\bibitem[{{Oka} {et~al.}(2001){Oka}, {Hasegawa}, {Sato}, {Tsuboi}, {Miyazaki},
  \& {Sugimoto}}]{oka01}
{Oka}, T., {Hasegawa}, T., {Sato}, F., {Tsuboi}, M., {Miyazaki}, A., \&
  {Sugimoto}, M. 2001, \apj, 562, 348

\bibitem[{{Pahre} {et~al.}(2004){Pahre}, {Ashby}, {Fazio}, \&
  {Willner}}]{pahre04ch1}
{Pahre}, M.~A., {Ashby}, M.~L.~N., {Fazio}, G.~G., \& {Willner}, S.~P. 2004,
  \apjs, 154, 235

\bibitem[{{Pineda} {et~al.}(2009){Pineda}, {Rosolowsky}, \&
  {Goodman}}]{pineda09}
{Pineda}, J.~E., {Rosolowsky}, E.~W., \& {Goodman}, A.~A. 2009, \apjl, 699,
  L134

\bibitem[{{Portegies Zwart} {et~al.}(2010){Portegies Zwart}, {McMillan}, \&
  {Gieles}}]{zwart10}
{Portegies Zwart}, S.~F., {McMillan}, S.~L.~W., \& {Gieles}, M. 2010, \araa,
  48, 431

\bibitem[{{Renaud} {et~al.}(2009){Renaud}, {Boily}, {Naab}, \&
  {Theis}}]{renaud09}
{Renaud}, F., {Boily}, C.~M., {Naab}, T., \& {Theis}, C. 2009, \apj, 706, 67

\bibitem[{{Sault} {et~al.}(1995){Sault}, {Teuben}, \& {Wright}}]{sault95}
{Sault}, R.~J., {Teuben}, P.~J., \& {Wright}, M.~C.~H. 1995, in Astronomical
  Society of the Pacific Conference Series, Vol.~77, Astronomical Data Analysis
  Software and Systems IV, ed. {R.~A.~Shaw, H.~E.~Payne, \& J.~J.~E.~Hayes},
  433

\bibitem[{{Sawada} {et~al.}(2001){Sawada}, {Hasegawa}, {Handa}, {Morino},
  {Oka}, {Booth}, {Bronfman}, {Hayashi}, {Luna Castellanos}, {Nyman},
  {Sakamoto}, {Seta}, {Shaver}, {Sorai}, \& {Usuda}}]{sawada01}
{Sawada}, T., {et~al.} 2001, \apjs, 136, 189

\bibitem[{{Schulz} {et~al.}(2007){Schulz}, {Henkel}, {Muders}, {Mao},
  {R{\"o}llig}, \& {Mauersberger}}]{schulz07}
{Schulz}, A., {Henkel}, C., {Muders}, D., {Mao}, R.~Q., {R{\"o}llig}, M., \&
  {Mauersberger}, R. 2007, \aap, 466, 467

\bibitem[{{Schweizer} {et~al.}(2008){Schweizer}, {Burns}, {Madore}, {Mager},
  {Phillips}, {Freedman}, {Boldt}, {Contreras}, {Folatelli}, {Gonz{\'a}lez},
  {Hamuy}, {Krzeminski}, {Morrell}, {Persson}, {Roth}, \&
  {Stritzinger}}]{schweizer08}
{Schweizer}, F., {et~al.} 2008, \aj, 136, 1482

\bibitem[{{Sheth} {et~al.}(2008){Sheth}, {Vogel}, {Wilson}, \&
  {Dame}}]{sheth08}
{Sheth}, K., {Vogel}, S.~N., {Wilson}, C.~D., \& {Dame}, T.~M. 2008, \apj, 675,
  330

\bibitem[{{Shu}(1977)}]{shu77}
{Shu}, F.~H. 1977, \apj, 214, 488

\bibitem[{{Snijders} {et~al.}(2007){Snijders}, {Kewley}, \& {van der
  Werf}}]{snijders07}
{Snijders}, L., {Kewley}, L.~J., \& {van der Werf}, P.~P. 2007, \apj, 669, 269

\bibitem[{{Solomon} {et~al.}(1992){Solomon}, {Downes}, \&
  {Radford}}]{solomon92}
{Solomon}, P.~M., {Downes}, D., \& {Radford}, S.~J.~E. 1992, \apjl, 398, L29

\bibitem[{{Solomon} {et~al.}(1997){Solomon}, {Downes}, {Radford}, \&
  {Barrett}}]{solomon97}
{Solomon}, P.~M., {Downes}, D., {Radford}, S.~J.~E., \& {Barrett}, J.~W. 1997,
  \apj, 478, 144

\bibitem[{{Solomon} {et~al.}(1987){Solomon}, {Rivolo}, {Barrett}, \&
  {Yahil}}]{solomon87}
{Solomon}, P.~M., {Rivolo}, A.~R., {Barrett}, J., \& {Yahil}, A. 1987, \apj,
  319, 730

\bibitem[{{Ueda} {et~al.}(2012){Ueda}, {Iono}, {Petitpas}, {Yun}, {Ho},
  {Kawabe}, {Mao}, {Mart{\'{\i}}n}, {Matsushita}, {Peck}, {Tamura}, {Wang},
  {Wang}, {Wilson}, \& {Zhang}}]{ueda11}
{Ueda}, J., {et~al.} 2012, \apj, 745, 65

\bibitem[{{van den Bergh}(2001)}]{vandenbergh01}
{van den Bergh}, S. 2001, \apjl, 559, L113

\bibitem[{{Wang} {et~al.}(2004){Wang}, {Fazio}, {Ashby}, {Huang}, {Pahre},
  {Smith}, {Willner}, {Forrest}, {Pipher}, \& {Surace}}]{wang04}
{Wang}, Z., {et~al.} 2004, \apjs, 154, 193

\bibitem[{{Whitmore}(2003)}]{whitmore03}
{Whitmore}, B.~C. 2003, in A Decade of Hubble Space Telescope Science, ed.
  {M.~Livio, K.~Noll, \& M.~Stiavelli}, 153--178

\bibitem[{{Whitmore} \& {Schweizer}(1995)}]{whitmore95}
{Whitmore}, B.~C., \& {Schweizer}, F. 1995, \aj, 109, 960

\bibitem[{{Whitmore} {et~al.}(1999){Whitmore}, {Zhang}, {Leitherer}, {Fall},
  {Schweizer}, \& {Miller}}]{whitmore99}
{Whitmore}, B.~C., {Zhang}, Q., {Leitherer}, C., {Fall}, S.~M., {Schweizer},
  F., \& {Miller}, B.~W. 1999, \aj, 118, 1551

\bibitem[{{Whitmore} {et~al.}(2010){Whitmore}, {Chandar}, {Schweizer},
  {Rothberg}, {Leitherer}, {Rieke}, {Rieke}, {Blair}, {Mengel}, \&
  {Alonso-Herrero}}]{whitmore10}
{Whitmore}, B.~C., {et~al.} 2010, \aj, 140, 75

\bibitem[{{Williams} {et~al.}(1994){Williams}, {de Geus}, \&
  {Blitz}}]{williams94}
{Williams}, J.~P., {de Geus}, E.~J., \& {Blitz}, L. 1994, \apj, 428, 693

\bibitem[{{Wilson} {et~al.}(2000){Wilson}, {Scoville}, {Madden}, \&
  {Charmandaris}}]{wilson00}
{Wilson}, C.~D., {Scoville}, N., {Madden}, S.~C., \& {Charmandaris}, V. 2000,
  \apj, 542, 120

\bibitem[{{Wilson} {et~al.}(2003){Wilson}, {Scoville}, {Madden}, \&
  {Charmandaris}}]{wilson03}
---. 2003, \apj, 599, 1049

\bibitem[{{Wu} {et~al.}(2005){Wu}, {Evans}, {Gao}, {Solomon}, {Shirley}, \&
  {Vanden Bout}}]{wu05}
{Wu}, J., {Evans}, II, N.~J., {Gao}, Y., {Solomon}, P.~M., {Shirley}, Y.~L., \&
  {Vanden Bout}, P.~A. 2005, \apjl, 635, L173

\bibitem[{{Zepf} {et~al.}(1999){Zepf}, {Ashman}, {English}, {Freeman}, \&
  {Sharples}}]{zepf99}
{Zepf}, S.~E., {Ashman}, K.~M., {English}, J., {Freeman}, K.~C., \& {Sharples},
  R.~M. 1999, \aj, 118, 752

\bibitem[{{Zezas} \& {Fabbiano}(2002)}]{zazas02b}
{Zezas}, A., \& {Fabbiano}, G. 2002, \apj, 577, 726

\bibitem[{{Zezas} {et~al.}(2002){Zezas}, {Fabbiano}, {Rots}, \&
  {Murray}}]{zazas02a}
{Zezas}, A., {Fabbiano}, G., {Rots}, A.~H., \& {Murray}, S.~S. 2002, \apj, 577,
  710

\bibitem[{{Zhang} {et~al.}(2010){Zhang}, {Gao}, \& {Kong}}]{zhang10}
{Zhang}, H.-X., {Gao}, Y., \& {Kong}, X. 2010, \mnras, 401, 1839

\bibitem[{{Zhang} \& {Fall}(1999)}]{zhang99}
{Zhang}, Q., \& {Fall}, S.~M. 1999, \apjl, 527, L81

\bibitem[{{Zhang} {et~al.}(2001){Zhang}, {Fall}, \& {Whitmore}}]{zhang01}
{Zhang}, Q., {Fall}, S.~M., \& {Whitmore}, B.~C. 2001, \apj, 561, 727

\bibitem[{{Zhu} {et~al.}(2003){Zhu}, {Seaquist}, \& {Kuno}}]{zhu03}
{Zhu}, M., {Seaquist}, E.~R., \& {Kuno}, N. 2003, \apj, 588, 243

\end{thebibliography}
\end{document}